\documentclass[longauth]{aa}

\newcommand{\Msun}{\ensuremath{{\rm M}_\odot}}                  
\newcommand{\Rsun}{\ensuremath{{\rm R}_\odot}}                  
\newcommand{\Teff}{\ensuremath{{\rm T}_{\rm eff}}}                      
\newcommand{\logg}{\ensuremath{\log g}}                           
\newcommand{\degrees}{\ensuremath{^\circ}}                        

\newcommand{\EBV}{\ensuremath{\textrm{E}\left(\textrm{B}-\textrm{V}\right)}}
\newcommand{\Ks}{\ensuremath{\textrm{K}_{\,\textrm{s}}}}
\newcommand{\vtci}{\ensuremath{\textrm{v}_{\textrm{\,TCI}}}}
\newcommand{\rtci}{\ensuremath{\textrm{r}_{\textrm{\,TCI}}}}
\newcommand{\dvtci}{\ensuremath{\Delta \textrm{v}_{\textrm{\,TCI}}}}
\newcommand{\drtci}{\ensuremath{\Delta \textrm{r}_{\textrm{\,TCI}}}}

\usepackage{graphicx}
\usepackage{txfonts}
\usepackage{natbib}

\usepackage{color}
\usepackage{epstopdf}
\usepackage{longtable}

\usepackage{lscape}

\bibpunct{(}{)}{;}{a}{}{,} 

\begin{document}

\title{High-resolution Imaging of Transiting Extrasolar Planetary systems (HITEP).}
	\subtitle{II. Lucky Imaging results from 2015 and 2016\thanks{Based on data collected by the MiNDSTEp consortium using the Danish 1.54 m telescope at the ESO La Silla observatory.}\ \thanks{Tables 2, 3, 4, 9, and 10 are only available in electronic form at the CDS via anonymous ftp to cdsarc.u-strasbg.fr (130.79.128.5) or via http://cdsweb.u-strasbg.fr/cgi-bin/qcat?J/A+A/}}
\author{D.\ F.\ Evans \inst{\ref{keele}}
	\and J.\ Southworth \inst{\ref{keele}}
	\and B.\ Smalley \inst{\ref{keele}}
	\and U.\ G.\ J{\o}rgensen \inst{\ref{nbicsp}}
	\and M.\ Dominik \inst{\ref{sta}}
	\and M.\ I.\ Andersen \inst{\ref{nbidcc}}
	\and V.\ Bozza \inst{\ref{salerno}, \ref{infn}}
	\and D.\ M.\ Bramich \inst{\ref{noaffil}}
	\and M.\ J.\ Burgdorf \inst{\ref{hamburg}}
	\and S.\ Ciceri \inst{\ref{mpia}}
	\and G.\ D'Ago \inst{\ref{inafnaples}}
	\and R.\ Figuera Jaimes \inst{\ref{sta}, \ref{esog}}
	\and S.-H.\ Gu \inst{\ref{yunnan}, \ref{kunming}}
	\and T.\ C.\ Hinse \inst{\ref{kasi}}
	\and Th.\ Henning \inst{\ref{mpia}}
	\and M.\ Hundertmark \inst{\ref{heidelberg}}
	\and N.\ Kains \inst{\ref{stsci}}
	\and E.\ Kerins \inst{\ref{jodrell}}
	\and H.\ Korhonen \inst{\ref{nbidcc}, \ref{finca}}
	\and R.\ Kokotanekova \inst{\ref{mpss}, \ref{openpss}}
	\and M.\ Kuffmeier \inst{\ref{nbicsp}}
	\and P.\ Longa-Pe\~{n}a \inst{\ref{antofa}}
	\and L.\ Mancini \inst{\ref{rome}, \ref{mpia}, \ref{inafturin}}
	\and J.\ MacKenzie \inst{\ref{nbicsp}}
	\and A.\ Popovas \inst{\ref{nbicsp}}
	\and M.\ Rabus \inst{\ref{puc}, \ref{mpia}}
	\and S.\ Rahvar \inst{\ref{sharif}}
	\and S.\ Sajadian \inst{\ref{isfahan}}
	\and C.\ Snodgrass \inst{\ref{openpss}}
	\and J.\ Skottfelt \inst{\ref{openpss}} 
	\and J.\ Surdej \inst{\ref{liege}}
	\and R.\ Tronsgaard \inst{\ref{aarhus}}
	\and E.\ Unda-Sanzana \inst{\ref{antofa}}
	\and C.\ von Essen \inst{\ref{aarhus}}
	\and Yi-Bo\ Wang \inst{\ref{yunnan}, \ref{kunming}}
	\and O.\ Wertz \inst{\ref{bonn}}
	}

\institute{
	Astrophysics Group, Keele University, Staffordshire, ST5 5BG, UK \email{d.f.evans@keele.ac.uk} \label{keele}
	\and Niels Bohr Institute \& Centre for Star and Planet Formation, University of Copenhagen {\O}ster Voldgade 5, 1350 - Copenhagen, Denmark \label{nbicsp}
	\and Centre for Exoplanet Science, SUPA School of Physics \& Astronomy, University of St Andrews, North Haugh, St Andrews, KY16 9SS \label{sta}
	\and Dark Cosmology Centre, Niels Bohr Institute, University of Copenhagen, Juliane Maries Vej 30, DK-2100 Copenhagen Ø \label{nbidcc}
	\and Dipartimento di Fisica "E.R. Caianiello", Universit{\`a} di Salerno, Via Giovanni Paolo II 132, 84084, Fisciano, Italy \label{salerno}
	\and Istituto Nazionale di Fisica Nucleare, Sezione di Napoli, Napoli, Italy \label{infn}
	\and No affiliation \label{noaffil}
	\and Universität Hamburg, Faculty of Mathematics, Informatics and Natural Sciences, Department of Earth Sciences, Meteorological Institute, Bundesstraße 55, 20146 Hamburg, Germany \label{hamburg}
	\and Max Planck Institute for Astronomy, K{\"o}nigstuhl 17, 69117 Heidelberg, Germany \label{mpia}
	\and INAF-Osservatorio Astronomico di Capodimonte, Salita Moiariello 16, 80131, Napoli, Italy \label{inafnaples}
	\and European Southern Observatory, Karl-Schwarzschild Stra\ss{}e 2, 85748 Garching bei M\"{u}nchen, Germany \label{esog}
	\and Yunnan Observatories, Chinese Academy of Sciences, Kunming 650011, China \label{yunnan}
	\and Key Laboratory for the Structure and Evolution of Celestial Objects, Chinese Academy of Sciences, Kunming 650011, China \label{kunming}
	\and Korea Astronomy \& Space Science Institute, 776 Daedukdae-ro, Yuseong-gu, 305-348 Daejeon, Republic of Korea \label{kasi}
	\and Astronomisches Rechen-Institut, Zentrum f\"{u}r Astronomie der Universit\"{a}t Heidelberg, M\"{o}nchhofstr. 12-14, 69120 Heidelberg, Germany \label{heidelberg}
	\and Space Telescope Science Institute, 3700 San Martin Drive, Baltimore, MD 21218, United States of America \label{stsci}
	\and Jodrell Bank Centre for Astrophysics, School of Physics and Astronomy, University of Manchester, Oxford Road, Manchester M13 9PL, UK \label{jodrell}
	\and Finnish Centre for Astronomy with ESO (FINCA), V{\"a}is{\"a}l{\"a}ntie 20, FI-21500 Piikki{\"o}, Finland \label{finca}
	\and Max Planck Institute for Solar System Research, Justus-von-Liebig-Weg 3, 37077 G{\"o}ttingen, Germany \label{mpss}
	\and School of Physical Sciences, Faculty of Science, Technology, Engineering and Mathematics, The Open University, Walton Hall, Milton Keynes, MK7 6AA, UK \label{openpss}
	\and Unidad de Astronom{\'{\i}}a, Fac. de Ciencias B{\'a}sicas, Universidad de Antofagasta, Avda. U. de Antofagasta 02800, Antofagasta, Chile \label{antofa}
	\and Department of Physics, University of Rome Tor Vergata, Via della Ricerca Scientifica 1, 00133 -- Roma, Italy \label{rome}
	\and INAF -- Astrophysical Observatory of Turin, Via Osservatorio 20, 10025 -- Pino Torinese, Italy \label{inafturin}
	\and Instituto de Astrof\'isica, Facultad de F\'isica, Pontificia Universidad Cat\'olica de Chile, Av. Vicu\~na Mackenna 4860, 7820436 Macul, Santiago, Chile \label{puc}
	\and Department of Physics, Sharif University of Technology, PO Box 11155-9161 Tehran, Iran \label{sharif}
	\and Department of Physics, Isfahan University of Technology, Isfahan 84156-83111, Iran \label{isfahan}
	\and Institut d'Astrophysique et de G\'eophysique, All\'ee du 6 Ao\^ut 19c, Sart Tilman, B\^at. B5c, 4000 Li\`ege, Belgium \label{liege} 
	\and Stellar Astrophysics Centre, Department of Physics and Astronomy, Aarhus University, Ny Munkegade 120, DK-8000 Aarhus C, Denmark \label{aarhus}
	\and Argelander-Institut f\"ur Astronomie, Auf dem Hügel, 71, 53121 Bonn, Germany \label{bonn}
	}

\date{Received -; accepted -}

\abstract{The formation and dynamical history of hot Jupiters is currently debated, with wide stellar binaries having been suggested as a potential formation pathway. Additionally, contaminating light from both binary companions and unassociated stars can significantly bias the results of planet characterisation studies, but can be corrected for if the properties of the contaminating star are known. }
{We search for binary companions to known transiting exoplanet host stars, in order to determine the multiplicity properties of hot Jupiter host stars. We also search for and characterise unassociated stars along the line of sight, allowing photometric and spectroscopic observations of the planetary system to be corrected for contaminating light. }
{We analyse lucky imaging observations of 97 Southern hemisphere exoplanet host stars, using the Two Colour Instrument on the Danish 1.54m telescope. For each detected companion star, we determine flux ratios relative to the planet host star in two passbands, and measure the relative position of the companion. The probability of each companion being physically associated was determined using our two-colour photometry. }
{A catalogue of close companion stars is presented, including flux ratios, position measurements, and estimated companion star temperature. For companions that are potential binary companions, we review archival and catalogue data for further evidence. For WASP-77AB and WASP-85AB, we combine our data with historical measurements to determine the binary orbits, showing them to be moderately eccentric and inclined to the line of sight (and hence planetary orbital axis). Combining our survey with the similar Friends of Hot Jupiters survey, we conclude that known hot Jupiter host stars show a deficit of high mass stellar companions compared to the field star population; however, this may be a result of the biases in detection and target selection by ground-based surveys. }
{}


\maketitle


\section{Introduction}

\label{sec:intro}

In the last two decades, many exoplanets have been discovered in orbital configurations that provide challenges for planet formation theory. One class of planets which has generated much interest are the hot Jupiters, a group of gas giants living extremely close to their host stars, with orbital periods of less than 10 days, with some examples having periods below a day. At the time of their discovery, planet formation theory suggested that gas giants form far from their host stars, where temperatures in the protoplanetary disc were low enough for volatiles -- such as water and methane -- to condense, permitting these planets to accrete sufficient mass to hold onto a gaseous envelope (e.g. \citealt{1996Icar..124...62P, 1995Sci...267..360B}). Theoretical models aiming to explain the formation of these planets generally suggest that the planet was indeed formed far from its star, but has since migrated inwards through some mechanism. However, some studies have also suggested that hot Jupiters may in fact be able to form in-situ \citep{2000Icar..143....2B, 2016ApJ...817L..17B, 2016ApJ...829..114B}.

Migration mechanisms generally fall into two broad categories: disc-planet interactions, which result in the planet losing angular momentum and energy to the protoplanetary disc and spiralling inwards towards its star (e.g. \citealt{1996Natur.380..606L, 2012ARA&A..50..211K}), and high eccentricity migration, where the planet is forced to a high eccentricity by dynamical interactions, and loses angular momentum through tidal or magnetic interactions with its host star. For the latter scenario, it has been suggested that the dynamical interactions could be between multiple planets in the same system, either with violent scattering events \citep{1996Sci...274..954R, 1996Natur.384..619W} or slower, long-term evolution \citep{2011Natur.473..187N, 2011ApJ...735..109W}. Alternatively, if the planet exists in a wide stellar binary, it is possible for the outer stellar companion to force the planet to high eccentricities via the Lidov-Kozai mechanism, which involves the exchange of angular momentum between the binary and planetary orbits \citep{2003ApJ...589..605W, 2007ApJ...669.1298F, 2012ApJ...754L..36N}.

If wide binaries are indeed responsible for hot Jupiter formation, it would therefore be expected that many hot Jupiters would be found in binary systems. It is also of interest to determine the stellar environments in which hot Jupiters do not exist. For example, it is expected that binary companions that are too close will inhibit planet formation \citep{2015ApJ...799..147J}, with some observational evidence supporting this theory \citep{2016AJ....152....8K}. It has also been suggested that wide binary companions, 1000au or so from the planet host star, could also have an influence on the planet at later stages if the binary orbit is modified by the galactic tide and close stellar encounters \citep{2013Natur.493..381K}. High resolution imaging allows the detection and confirmation of wide stellar binaries, through the use of multi-colour photometry and common proper motion analysis.

High resolution imaging is also able to discover physically unassociated stars in the vicinity of suspected transiting exoplanet host stars. Transit surveys, especially those operating from the ground, suffer from a high rate of false positives, with grazing, blended, or high mass ratio eclipsing binaries being able to mimic the signals of a planet \citep{2003ApJ...593L.125B, 2017MNRAS.465.3379G}. However, space-based missions are also liable to false positives, due to faint targets or small photometric/spectroscopic signals making ground-based follow-up observations difficult, as was recently shown in the case of three claimed transiting planets from the K2 mission \citep{2017arXiv170708007C}.

When the planetary signal is real, it is still important to consider the effect of other stars. Contaminating light can significantly alter the derived bulk properties of a planet \citep{2011ApJS..197....3B, 2016ApJ...833L..19E}. Additionally, a nearby star dissimilar to the planet host star can alter or mimic various features that are expected to be seen in planetary atmospheres, with cool, red stars producing slopes towards the blue that resemble Rayleigh scattering \citep{2016MNRAS.463...37S}.

In this paper, we present the continuation of our lucky imaging survey of transiting exoplanet host stars in the Southern hemisphere. In \citet{HITEP1}, hereafter Paper I, we presented lucky imaging observations of 101 planet host stars, detecting 51 companions located within 5 arcseconds of a target star. The current paper covers lucky imaging observations of 97 targets, including re-observations of several interesting objects, with two colour photometry allowing us to assess the likelihood of each detected companion star being bound. 

\begin{table}
	\caption{\label{tab:quickobs} Summary of transiting exoplanet systems considered in this paper. Systems with potential or confirmed binary companions are shown in bold (see Section~\ref{sec:results}). }
	\centering
	\begin{tabular}{l l l l} \hline \hline
		\multicolumn{4}{c}{Systems studied in this work} \\
		\hline
		{\bf CoRoT-02} & HATS-16        & WASP-25        & {\bf WASP-87}  \\
		CoRoT-04       & HATS-17        & {\bf WASP-26}  & {\bf WASP-94}  \\
		CoRoT-07       & HATS-18        & WASP-28        & WASP-97        \\
		{\bf CoRoT-22} & HATS-25        & WASP-31        & WASP-101       \\
		{\bf CoRoT-24} & HATS-26        & WASP-32        & {\bf WASP-104} \\
		{\bf CoRoT-28} & HATS-27        & {\bf WASP-36}  & {\bf WASP-106} \\
		GJ 1132        & HATS-29        & WASP-37        & {\bf WASP-108} \\
		HAT-P-26       & {\bf K2-02}    & WASP-39        & {\bf WASP-109} \\
		HAT-P-27       & K2-03          & WASP-41        & WASP-110       \\
		{\bf HAT-P-30} & K2-10          & WASP-42        & {\bf WASP-111} \\
		{\bf HAT-P-35} & K2-19          & WASP-44        & WASP-112       \\
		{\bf HAT-P-41} & K2-21          & {\bf WASP-45}  & WASP-117       \\
		HAT-P-45       & K2-24          & {\bf WASP-49}  & {\bf WASP-121} \\
		HAT-P-46       & {\bf K2-27}    & {\bf WASP-54}  & WASP-122       \\
		{\bf HATS-01}  & K2-31          & {\bf WASP-55}  & {\bf WASP-123} \\
		HATS-02        & {\bf K2-38}    & WASP-64        & WASP-124       \\
		HATS-03        & K2-39          & {\bf WASP-66}  & {\bf WASP-129} \\
		HATS-04        & K2-44          & WASP-67        & WASP-130       \\
		HATS-07        & {\bf KELT-15}  & {\bf WASP-68}  & WASP-131       \\
		HATS-09        & {\bf WASP-08}  & {\bf WASP-70}  & WASP-132       \\
		{\bf HATS-10}  & WASP-16        & WASP-74        & {\bf WASP-133} \\
		HATS-11        & WASP-17        & {\bf WASP-77}  & WASP-157       \\
		{\bf HATS-12}  & {\bf WASP-19}  & WASP-80        &                \\
		HATS-13        & WASP-23        & WASP-83        &                \\
		{\bf HATS-14}  & WASP-24        & {\bf WASP-85}  &                \\
		\hline
	\end{tabular}
\end{table}

\section{Observations}

Lucky imaging observations of 97 targets were carried out using the Two Colour Instrument (TCI) on the Danish 1.54m telescope, located at La Silla, Chile. Following the observing strategy in Paper I, our targets are all publicly announced transiting exoplanet host stars, taken from the \texttt{TEPCat}\footnote{http://www.astro.keele.ac.uk/jkt/tepcat/} database of well-studied transiting extrasolar planets \citep{2011MNRAS.417.2166S}. The observations reported in this work were taken during the periods April-September 2015 and April-September 2016. Targets were generally observed once in each year, with new systems being added to our target lists as they were published. Targets for which no companions were detected in our 2014 and 2015 observations, or for which companions were determined to be physically unassociated, were not re-observed in 2016. A summary of targets observed as part of this work is given in Table~\ref{tab:quickobs}, with further details of individual observations given in Tables~\ref{tab:obslist15}~and~\ref{tab:obslist16}, which cover 2015 and 2016 respectively.

The TCI is a lucky imager designed for simultaneous two-colour photometry, using Electron Multiplying CCD (EMCCD) detectors, and is described in detail by \citet{2015A&A...574A..54S}. Light is split between two cameras by a dichroic, with the `red' camera receiving light redwards of 655nm, and the `visual' camera receiving light between 466nm and 655nm, with a second dichroic sending light bluewards of 466nm to a focus system. No filters are fitted to the system, and so the passband for each camera is defined only by the dichroics and the sensitivity of the EMCCD chips. We refer to the passband of the visual and red cameras as $\textrm{v}_{\textrm{\,TCI}}$ and $\textrm{r}_{\textrm{\,TCI}}$ respectively. The designed passband of the red camera is similar to the combination of the SDSS $i'$ and $z'$ filters, or a wider version of the Cousins $I$ filter. The visual camera has a central wavelength that is located between the SDSS $g'$ and $r'$ filters, similar to the Johnson $V$ filter's central wavelength, but has a significantly different response to any of the mentioned filters. Both detectors consist of a 512$\times$512 pixel array with a pixel scale of approximately $\sim$0.09 arcsec/pixel, giving a field of view of approximately $45\arcsec \times 45\arcsec$.

The target systems were selected by brightness ($9\leq V \leq 15$), with exposure times chosen to match this brightness, ranging from 60s to 900s. For most targets, the default electron multiplication gain of 300 e$^{-}$/photon was used, but targets brighter than $V=10.5$ were observed with a lower gain of 100 e$^{-}$/photon, in order to keep counts within the range of the analogue to digital converter. Observations were performed at an airmass of $1.5$ or less, and in seeing conditions of $0.6\arcsec$ or better. Targets were observed simultaneously with the visual and red cameras, with the exception of the observation of WASP-104 on the night of 2015-06-04, when a technical problem prevented data being taken with the visual camera.

\begin{table*}
		\caption{\label{tab:obslist15} Observations from 2015 analysed in this work. The ``1\% FWHM'' column lists the FWHM of the target star in the best 1\% of lucky imaging exposures.  The ``Detection Limit'' columns indicate the calculated detection limits in $\rtci$, determined using the method discussed in Section~\ref{sect:DetLimits}.  The full version of table is available in electronic form at the CDS. }
		\centering
		\begin{tabular}{l c c c c c c c c c c }
		\hline \hline 
		&      &               &                & \multicolumn{7}{c}{Detection limit (mag)} \\
		Target & Date & Exp. time (s) & 1\% FWHM (\arcsec) & 0.5\arcsec & 0.8\arcsec & 1.5\arcsec & 2.5\arcsec & 5.0\arcsec & 10.0\arcsec & 20.0\arcsec \\
		\hline
		CoRoT-22 & 2015-09-12 23:50 & 900 & 0.57 &  0.00 &  3.29 &  5.84 &  8.08 &  8.77 &  8.75 &  8.70 \\
		CoRoT-28 & 2015-09-12 23:32 & 900 & 0.53 &  0.00 &  4.06 &  6.53 &  8.77 &  9.48 &  9.52 &  9.47 \\
		HAT-P-27 & 2015-05-12 04:47 & 300 & 0.33 &  4.64 &  6.16 &  8.84 & 10.71 & 11.43 & 11.46 & 11.41 \\
		HAT-P-30 & 2015-04-28 01:30 & 60  & 0.52 &  2.32 &  3.84 &  6.25 &  8.43 &  9.64 &  9.76 &  9.74 \\
		\multicolumn{11}{c}{$\cdots$} \\
		\hline	
	\end{tabular}
\end{table*}

\begin{table*}
	\caption{\label{tab:obslist16} Observations from 2016 analysed in this work. The ``1\% FWHM'' column lists the FWHM of the target star in the best 1\% of lucky imaging exposures.  The ``Detection Limit'' columns indicate the calculated detection limits in $\rtci$, determined using the method discussed in Section~\ref{sect:DetLimits}. The full version of table is available in electronic form at the CDS. }
	\centering
	\begin{tabular}{l c c c c c c c c c c }
		\hline \hline 
		&      &               &                & \multicolumn{7}{c}{Detection limit (mag)} \\
		Target & Date & Exp. time (s) & 1\% FWHM (\arcsec) & 0.5\arcsec & 0.8\arcsec & 1.5\arcsec & 2.5\arcsec & 5.0\arcsec & 10.0\arcsec & 20.0\arcsec \\
		\hline
		CoRoT-02 & 2016-04-30 10:01 & 450 & 0.34 &  4.45 &  6.11 &  8.65 & 10.43 & 10.98 & 11.03 & 10.96 \\ 
		CoRoT-04 & 2016-04-21 00:51 & 900 & 0.70 &  0.00 &  2.90 &  5.43 &  7.56 &  8.19 &  8.19 &  8.14 \\ 
		CoRoT-07 & 2016-04-21 00:08 & 200 & 0.42 &  2.57 &  4.68 &  7.33 &  9.64 & 10.44 & 10.47 & 10.43 \\ 
		CoRoT-24 & 2016-04-21 00:35 & 850 & 0.86 &  0.00 &  0.00 &  4.49 &  6.53 &  6.65 &  6.65 &  6.60 \\ 
		\multicolumn{11}{c}{$\cdots$} \\
		\hline
	\end{tabular}
\end{table*}

\section{Data reduction and analysis}

In general, our reduction process follows the outline given in Paper I \citep{HITEP1}. No major changes have been made to the reduction pipeline, as described in \citet{2012A&A...542A..23H} and \citet{2015A&A...574A..54S}, which automatically reduces data from the TCI. The process includes bias and flat field corrections, subtraction of cosmic rays, tip-tilt correction, and measurement of the quality of the lucky imaging exposures. The ranked exposures are then divided into ten cuts, with the exposures in each cut being combined into a single stacked image. The percentage boundaries between each cut are: 1, 2, 5, 10, 20, 50, 90, 98, 99, 100. Therefore, the first cut contains the top 1\% of images ranked by quality, the second cut contains the next 1\% of images, and so on.

We apply the correction for smearing described in Paper I, which removes apparent smears across the image that appear when bright stars are present, thought to be caused by charge transfer inefficiency. We then run our star detection algorithm, again described in Paper I, which remains unchanged. In brief, we take the first seven cuts in quality (covering the best 90\% of the exposures), and convolve each cut with a Gaussian of standard deviation 4 pixels (FWHM 11.7 pixels). Each convolved image is then subtracted from the original, in order to filter out high frequency noise in the image. We then divide each image into a series of annuli centred on the target star, with each annulus having a width of 0.5 pixels. For each annulus, a sigma-clipped mean and standard deviation are calculated, and any pixels more than $1.7\sigma$ from the annulus mean are flagged as a candidate detection, this value having been chosen as the best compromise between false positives and missed detections. The detections from each of the seven cuts are then combined, with any groups of adjacent pixels marked as detections being grouped together into a single result.

Each detection is then verified by eye using an interactive program, with the main source of spurious detections being asymmetries of the target star's PSF. Additionally, in crowded fields it is possible for several overlapping stars to be combined by the pipeline into a single detection, or for the identified position of a faint star to be offset slightly due to an overlapping bright star; the user is able to correct the identifications in such cases.

The star detection algorithm was run only on the data from the red camera, as this passband is more sensitive to low mass stellar companions. Additionally, the visual images generally have a lower angular resolution; firstly, is not possible to independently focus the red and visual cameras, with the instrument focus being set based on the red camera images. As a result; secondly, due to the turbulent properties of the atmosphere, the effects of seeing are worse at shorter wavelengths \citep{1966JOSA...56.1372F}.

\subsection{Detection limits} \label{sect:DetLimits}

\begin{figure} 
	\includegraphics[width=\columnwidth,angle=0]{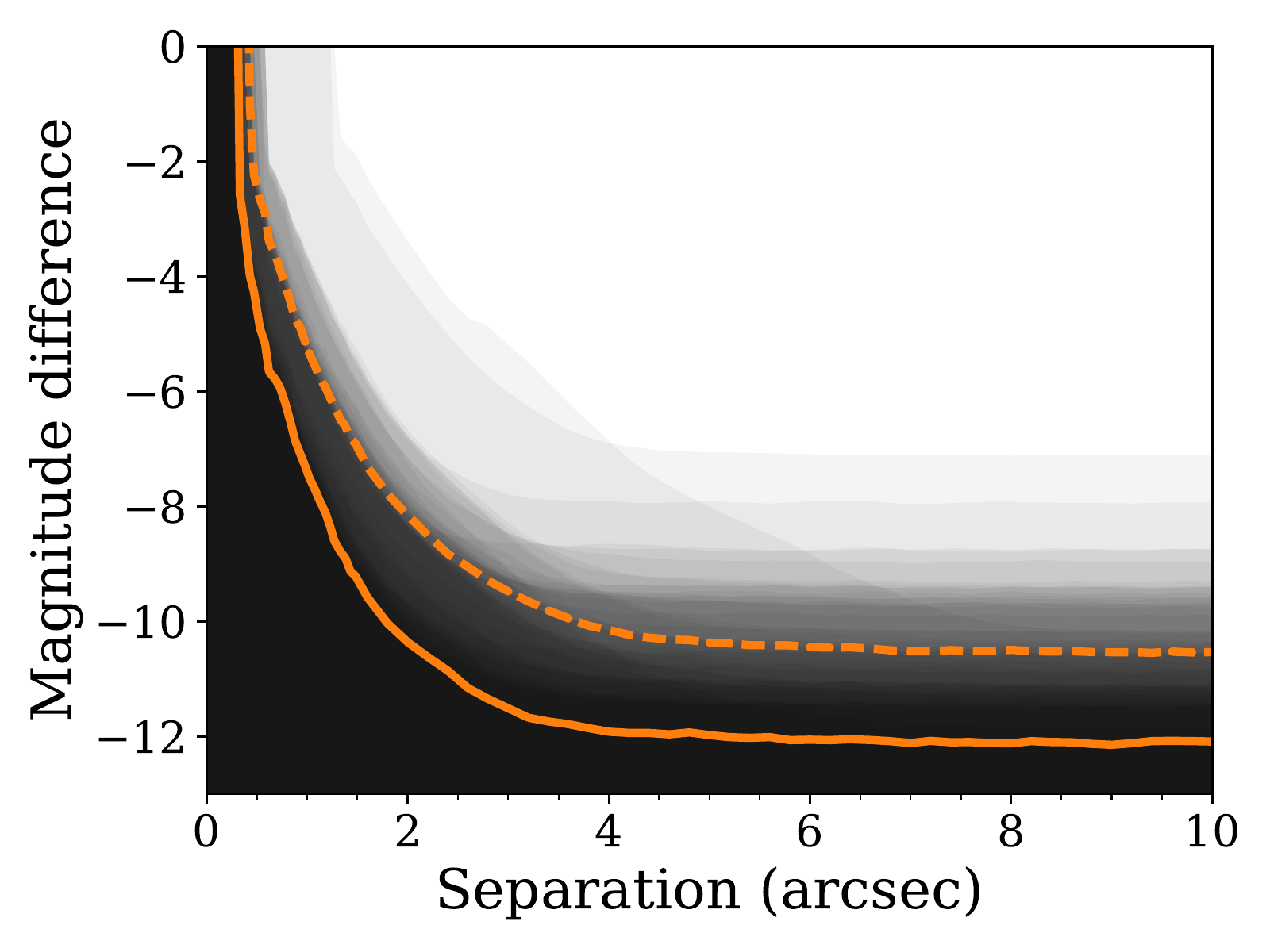}
	\caption{\label{fig:detectionlimits2015} The 5-sigma detection limits for our 2015 observations. The dashed line indicates the median sensitivity, whilst the solid line indicates the maximum sensitivity achieved for any observation. Shading on the figure indicates for what fraction of our observations a companion of a given separation and magnitude difference would be detectable. A region that is entirely black indicates that a companion would not be detected in any observation, whilst a region that is white indicates that a companion would be detected in all observations. All detection curves remain essentially flat beyond $10\arcsec$.}
\end{figure}

\begin{figure} 
	\includegraphics[width=\columnwidth,angle=0]{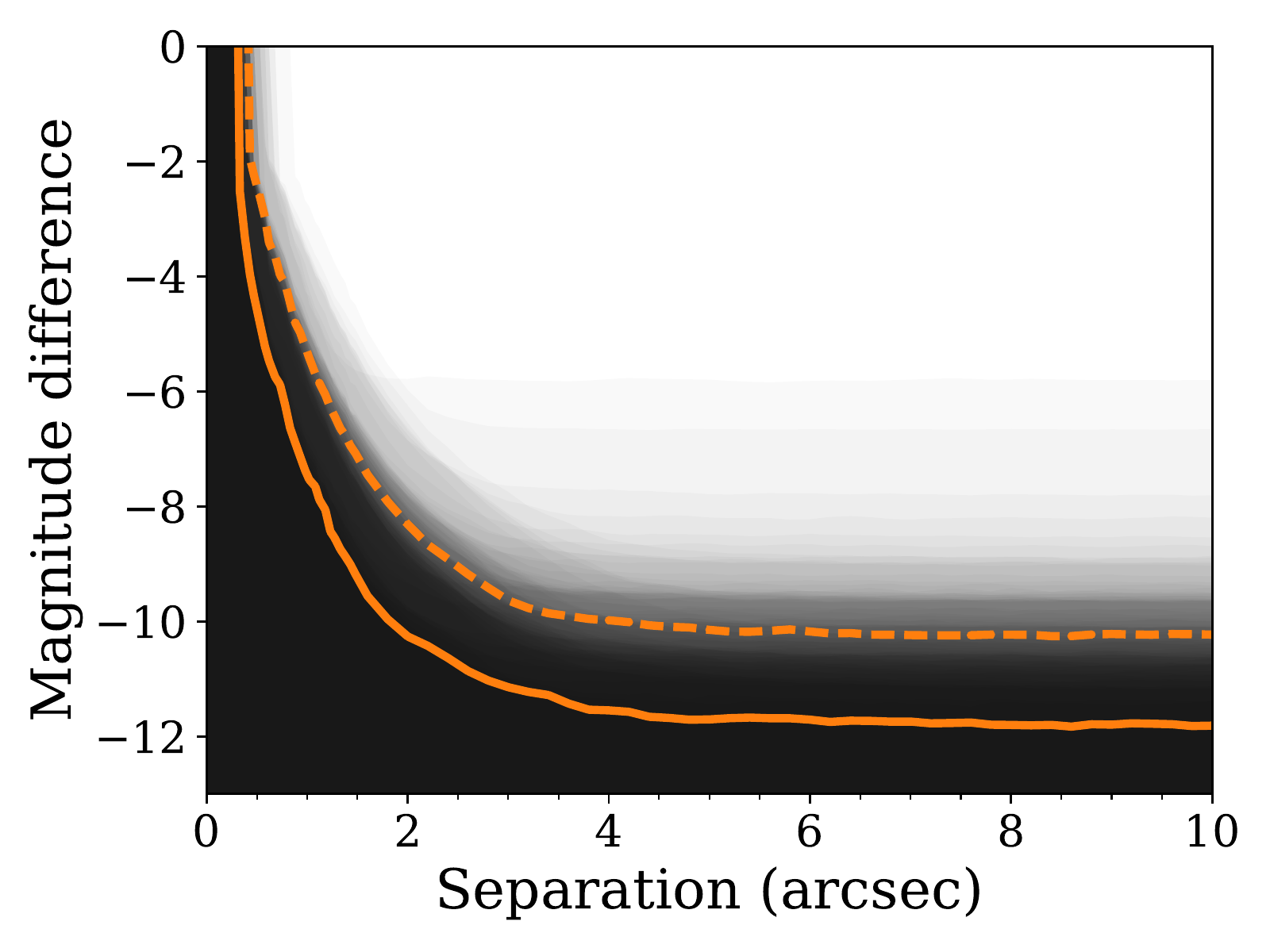}
	\caption{\label{fig:detectionlimits2016} The 5-sigma detection limits for our 2016 observations. See caption for Fig.~\ref{fig:detectionlimits2015}. }
\end{figure}

For a given pixel on an image, we determined the mean and standard deviation of the counts inside a box of $n\times n$ pixels, centred on the pixel in question, with $n$ being the FWHM of stars on the image. The number of counts corresponding to a $5\sigma$ excess were then calculated, giving the faintest detectable star. This process was repeated for each pixel on the image, with pixels then being binned by separation from the target star, and the median detection limit for each bin calculated. Limits were converted from counts to magnitudes using the counts on the central star within the same $n\times n$ box. This was repeated for each of the seven lucky imaging cuts, from which a final detection limit was compiled for each observation. Figs.~\ref{fig:detectionlimits2015}~and~\ref{fig:detectionlimits2016} illustrate the detection limits for all targets in 2015 and 2016 respectively. Additionally, the detection limits for each target are tabulated in Tables~\ref{tab:obslist15}~and~\ref{tab:obslist16}.

\subsection{Photometric calibration}

We adopt the set of theoretical colour indices presented in Paper I, which give the expected magnitude difference between $\textrm{v}_{\textrm{\,TCI}}$ and $\textrm{r}_{\textrm{\,TCI}}$, for main sequence stars and brown dwarfs between 2300K and 12000K. For stars below 7000K, the expected luminosity in both the visual and red was also presented, our calibration for radius (and hence luminosity) not being valid above this temperature. These two calibrations allowed us to estimate the temperature of a candidate companion star, and then to determine whether its brightness is consistent with being at the same distance as the target star, or instead that it is too bright or too faint compared to the target, indicating that it is closer or further away.

With the much more extensive two-colour dataset presented in this work, it became apparent that our previous temperature-radius calibration was not sufficiently accurate below $\sim3500$K, with known physical companions being fainter than expected. Comparison with the \citet{2015A&A...577A..42B} stellar models for an age of $1$Gyr shows that our radius calibration agrees to within a few percent down to $3500$K, but predicts much larger radii than the models at cooler temperatures. We therefore use the \citet{2015A&A...577A..42B} model radii below $4000$K, where the two sets of radii begin to deviate.

We adopt the previously calculated atmospheric extinction of $0.08\pm0.03$ mag/airmass for the $\left(\textrm{v}-\textrm{r}\right)_{\textrm{\,TCI}}$ colour index. In Paper I, we calibrated the predicted values of $\left(\textrm{v}-\textrm{r}\right)_{\textrm{\,TCI}}$ to the measured values. We present a new zero point offset based on our 2015 and 2016 data of $-0.405\pm0.011$ magnitudes -- i.e. a star with a predicted colour of $0.000$ would be measured as $0.405$.

Our analysis in Paper I showed a potential temperature-dependent colour offset, with hot standard stars having a slightly larger offset of $\sim0.50$ magnitudes, whilst cooler stars clustered around an offset of $\sim0.44$ magnitudes. The sample of target stars in this paper cover $3270-6600$K, and we find no correlation between temperature and colour offset. We find a scatter of $0.07$mag in our colour measurements, similar to the variation from $0.44$ to $0.50$ magnitudes. We attribute the scatter to atmospheric transparency variations, errors in the temperatures from which the expected colours were derived, and intrinsic physical variations due to effects such as metallicity.

\subsubsection{Host star distances} \label{sect:StarDists}

To convert a companion's angular separation from the host stars to projected separation, and hence begin to consider the orbital properties of the companion, we must know the distance to the host star. Many, but not all, hot Jupiter host stars have photometric or spectroscopic distance estimates published; however, the methodology used to derive these distances is varied, and in a number of cases, is simply not stated. With the release of the Tycho-Gaia astrometric solutions (TGAS, \citealt{2015A&A...574A.115M}), approximately half of the TEP hosts in surveys such as WASP and HAT now have measured parallaxes, but the uncertainties on the derived distances become significant beyond 300pc.

We therefore choose to derive our own distance estimates for our targets in a homogeneous fashion, using the K-band surface brightness-effective temperature relation presented in \citet{2004A&A...426..297K}, as previously done in Paper I. Stellar radii and effective temperatures were taken from the \texttt{TEPCat} database, using the version of the database from 2016-05-01\footnote{\texttt{TEPCat} is archived monthly, and the version used in this work can be found at: http://www.astro.keele.ac.uk/jkt/tepcat/2016may/allplanets-noerr.html}. K band magnitudes were taken from \texttt{2MASS} \Ks\ magnitudes \citep{2006AJ....131.1163S}, using the relation $\mathrm{K}=\Ks + 0.044$ \citep{2001AJ....121.2851C}.

To correct for interstellar extinction, we derived an initial distance estimate with no correction for extinction, from which we then obtained a value for \EBV\ from the three-dimensional dust map of \citet{2015ApJ...810...25G}, which provides measurements at several distances along a given line of sight. The extinction in \Ks was then calculated, using $\textrm{A}(\Ks) = 0.306\,\EBV$ \citep{2013MNRAS.430.2188Y}, and the extinction-corrected value of \Ks used to calculate a new distance estimate. This process was iterated until the change in calculated distance was less than 1\% between iterations.

The \EBV\ data provided by \citet{2015ApJ...810...25G} is tabulated in steps of $0.5$ in distance modulus, between which we interpolated linearly to get our values of \EBV. For stars closer than the nearest reddening measurement provided by the map, we assumed $\EBV=0$ at zero distance, and again interpolated linearly out to the nearest measurement. The map does not cover the entire sky, with a region towards the Southern pole currently being unmapped -- being away from the galactic plane, this is an area of relatively low reddening, and we therefore assumed extinction was negligible for targets in this area.

\begin{figure} 
	\includegraphics[width=\columnwidth,angle=0]{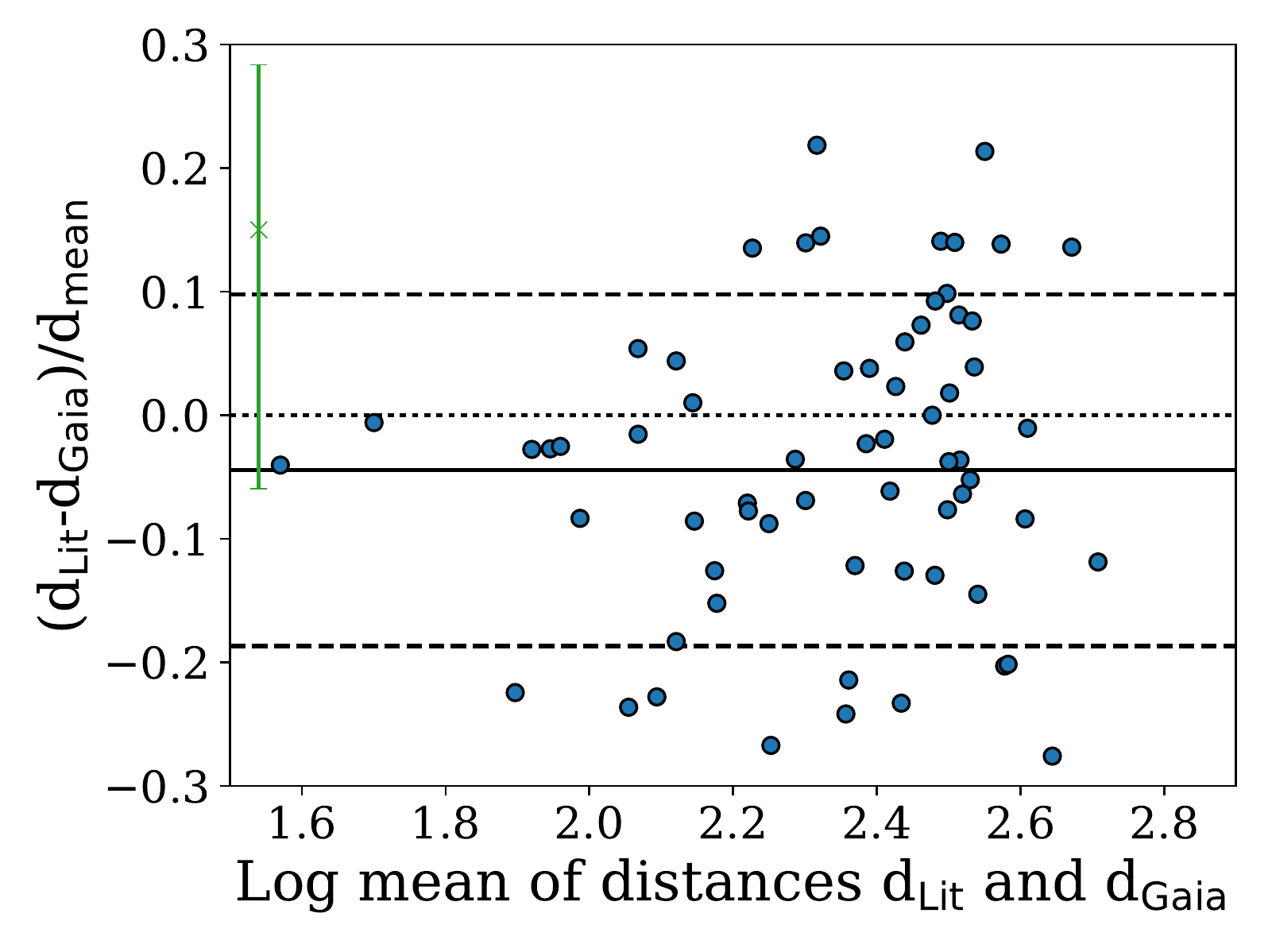}
	\caption{\label{fig:LiteraturevGAIA} A comparison between previously published distances $\rm d_{\rm Lit}$ and distances derived from the TGAS parallaxes $\rm d_{\rm Gaia}$. The figure shows all published HAT and WASP systems where both values are available, with each system plotted using the log of the mean of the two distance values, and the fractional difference between the two values. The green data point at top left illustrates the median error bar, with individual error bars excluded for clarity. The solid and dashed lines indicate the mean difference and RMS scatter between the two sets of data; the dotted line shows where points would fall if there were no systematic offset. We find that the exoplanet literature distances are systematically $4.4\pm1.5\%$ lower than the TGAS distances, with an RMS scatter of $14\%$ between the two sets of data. }
\end{figure}

\begin{figure} 
	\includegraphics[width=\columnwidth,angle=0]{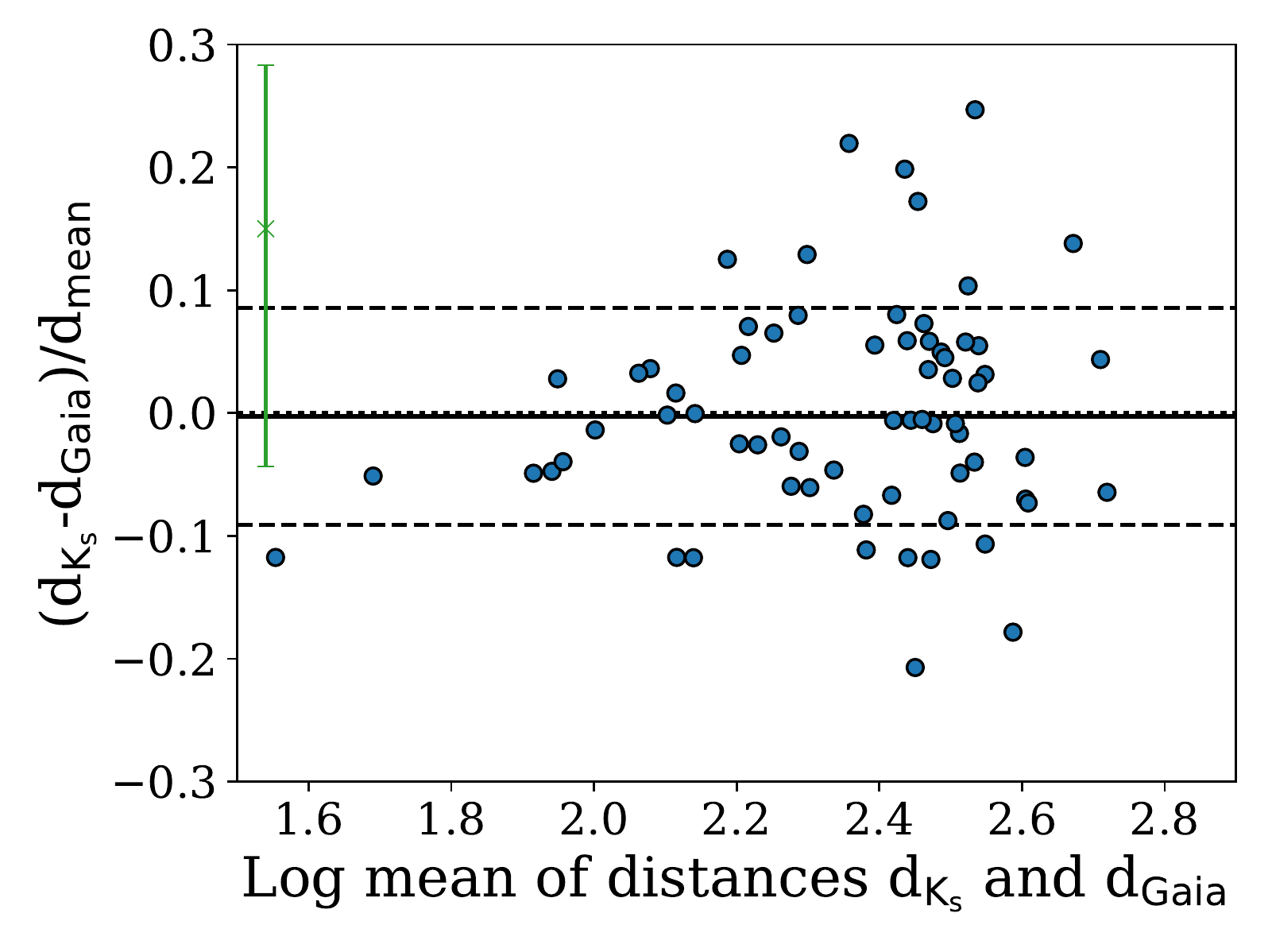}
	\caption{\label{fig:KbandvGAIA} As for Fig.~\ref{fig:LiteraturevGAIA}, comparing distances derived using the K-band surface brightness-effective temperature relation ${\rm d_{\rm Ks}}$ to those derived from the TGAS parallaxes $\rm d_{\rm Gaia}$. We find that our derived distances are in good agreement with the TGAS-derived distances, with our values being only $0.3\pm1.7\%$ higher on average, with an RMS scatter of $9\%$ between the two sets of data. }
\end{figure}

We compare both our distance values and those in the exoplanet literature for all HAT and WASP survey planets to those derived from the TGAS dataset in Figures~\ref{fig:LiteraturevGAIA}~and~\ref{fig:KbandvGAIA}. We adopt the distances and uncertainties calculated for the TGAS dataset by \citet{2016arXiv160907369A} using their anisotropic `Milky Way' prior on the distribution of stars, which shows the best agreement with Cepheid distances within 2kpc. We find that the literature distances are systematically lower than those found by TGAS, with a mean ratio of $d_{\rm Lit} / d_{\rm TGAS} = 0.956\pm0.015$ ($2.9\sigma$) with an RMS scatter of $14\%$. Our K-band derived distances show a negligible bias and a smaller scatter, with $d_{\rm Ks} / d_{\rm TGAS} = 0.997\pm0.018$ ($0.2\sigma$) and an RMS scatter of $9\%$. Our derived distances, and the values used to calculate them, can be found in Table~\ref{tab:dists}.

\begin{table*}
		
	\caption{\label{tab:dists} Assumed properties of our target stars, including distances estimated using the K-band surface brightness-effective temperature relation, and the values used to calculate the distances. Where no entry is given in the \EBV\ column, no extinction estimate was available, and it was assumed that the effect of extinction was negligible.  The full version of table is available in electronic form at the CDS. }
	\centering
	\begin{tabular}{l r@{\,$\pm$\,}l r@{\,$\pm$\,}l r@{\,$\pm$\,}l r@{\,$\pm$\,}l r@{\,$\pm$\,}l}
		\hline \hline 
		Target & \multicolumn{2}{c}{Radius\ (\Rsun)} & \multicolumn{2}{c}{\Teff\ (K$_{\rm s}$)} & \multicolumn{2}{c}{K (mag)} & \multicolumn{2}{c}{\EBV} & \multicolumn{2}{c}{Distance\ (pc)} \\
		\hline		
		CoRoT-02 &  0.90 &  0.09 & 5598 &  100 & 10.31 & 0.03 & 0.153 & 0.016 &  221.8 &  22.4 \\
		CoRoT-04 &  1.15 &  0.11 & 6190 &  100 & 12.29 & 0.03 & 0.125 & 0.011 &  756.8 &  76.7 \\
		CoRoT-07 &  0.82 &  0.08 & 5259 &  100 &  9.81 & 0.02 & 0.024 & 0.002 &  155.6 &  15.8 \\
		CoRoT-22 &  1.14 &  0.11 & 5939 &  100 & 11.99 & 0.03 & 0.175 & 0.014 &  630.2 &  63.9 \\
		\multicolumn{11}{c}{$\cdots$} \\
		\hline
	\end{tabular}
\end{table*}

\subsubsection{Interstellar reddening}

\begin{figure} 
	\includegraphics[width=\columnwidth,angle=0]{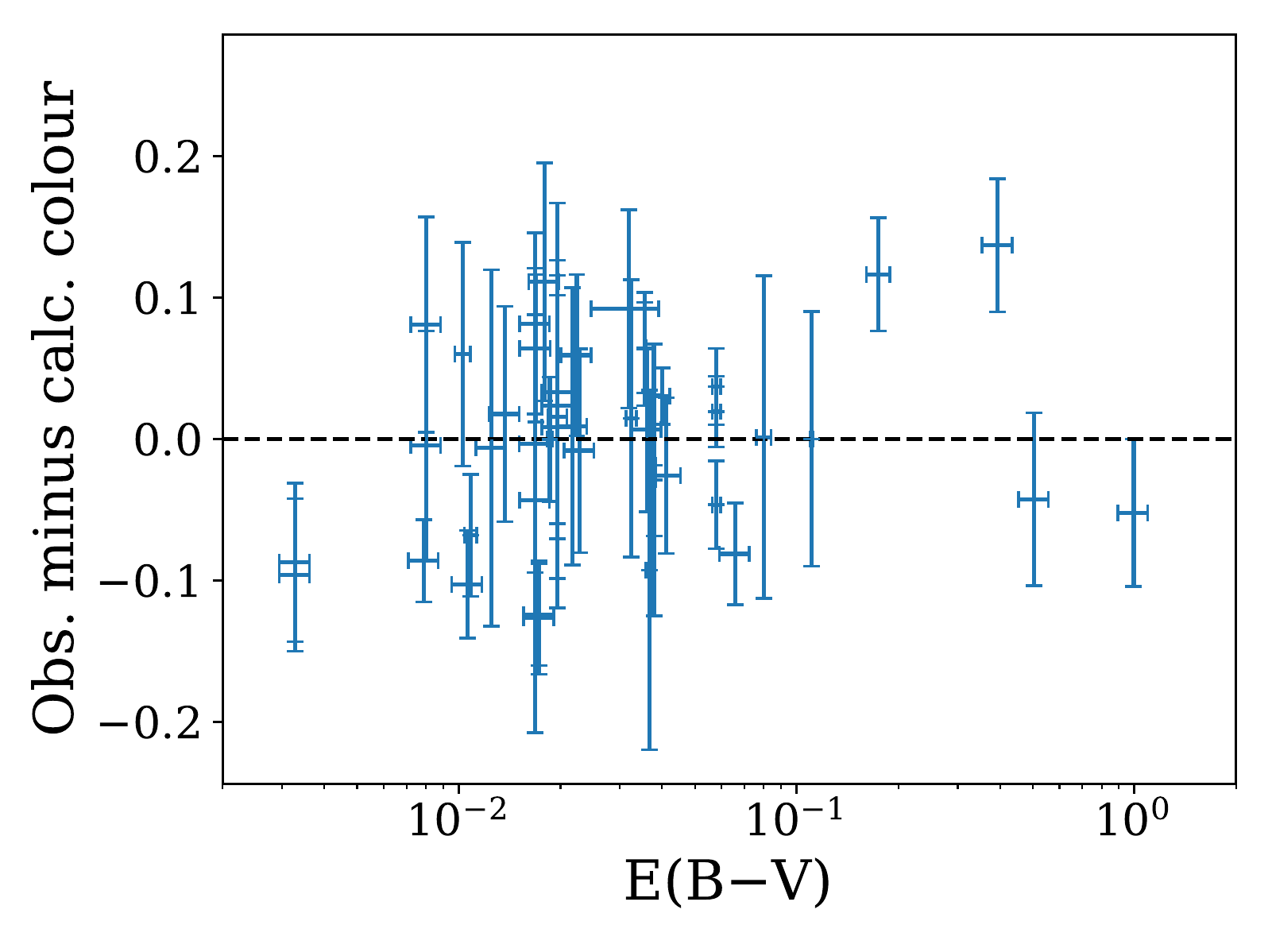}
	\caption{\label{fig:EBVs} The offset between the expected and measured colours of our target stars after correction for atmospheric effects and interstellar reddening, plotted against the expected \EBV\ for each target. Note that the x axis is logarithmic, as the majority of our targets have very low reddening values. All targets agree to within $\pm0.15$ dex, corresponding to  agreement within $\pm300$K. } 
\end{figure}

As part of our analysis of companions, temperatures are derived from our photometric data, using the colour-temperature calibrations in Paper I. To correct for interstellar reddening, which would bias our results towards lower temperatures, we use the \EBV\ values determined in the previous section, and listed in Table~\ref{tab:dists}. Between approximately $350$nm and $1000$nm, the relationship between interstellar extinction (measured in magnitudes) and wavelength is almost linear (e.g. \citealt{1999PASP..111...63F}). We therefore assumed that the TCI colour excess $\textrm{E}\left(\textrm{v}-\textrm{r}\right)_{\textrm{\,TCI}}$ is linearly related to the standard Johnson \EBV\ colour excess by some constant $X$,
\begin{equation}
\textrm{E}\left(\textrm{v}-\textrm{r}\right)_{\textrm{\,TCI}} = \left(\textrm{v}-\textrm{r}\right)_{\textrm{True}}-\left(\textrm{v}-\textrm{r}\right)_{\textrm{Obs}} = X \cdot \EBV
\end{equation}
We used the effective temperatures from \texttt{TEPCat} to calculate the expected intrinsic colours for our observed target stars. The colours for each target were then measured from our data, corrected for the zero-point offset and atmospheric extinction, and the colour excess (difference between observed and calculated) was derived. A least squares fit was then performed to derive a value for the reddening factor $X$, using the measured colour excesses and the value of \EBV\ that was derived when calculating the distance to each star. We calculated a value of $X=0.53\pm0.06$, and the agreement between expected and measured colours after applying this correction is shown in Fig.~\ref{fig:EBVs}.

\subsection{Companion colour analysis}

From our two-colour photometry and the colour indices presented in Paper I, we are able to estimate the temperature of detected companion stars. By comparing this to the literature temperature of the host star and assuming both are main sequence stars, we then calculate the expected flux ratio in both $\vtci$ and $\rtci$ if the target and companion form a physical binary and hence are at the same distance, using our temperature-radius calibration. Background stars will then appear fainter than expected, whilst foreground stars will be too bright, allowing us to assign a low probability of physical association to stars in a single epoch, without long-term proper motion monitoring. However, further evidence is required to reach definite conclusions. Chance alignments can result in stars with similar distances to appear close together, especially in cases where the target star is more distant or located in the plane of the galaxy. In addition, companions may not be main sequence stars -- for example, the K2-02 (HIP 116454) system includes a white dwarf companion \citep{2015ApJ...800...59V}, which would be classified as a distant background object from its blue colour.

No further observations of standard stars have occurred since Paper I, and the available data do not allow us to meaningfully constrain atmospheric extinction, which could bias our colour-based temperatures. To account for this, we assume that the colours of the target stars should match the colour predicted by the published spectroscopic temperatures listed in Table~\ref{tab:dists}, and use the offset between the measured and expected colours to determine the atmospheric extinction for each observation.

\subsection{Astrometric calibration}

To determine the detector scales and orientations for both the visual and red cameras, we compared TCI observations of the cores of globular clusters (GCs) to reference data, mainly derived from Hubble Space Telescope (HST) observations. The GC observations were obtained as part of a separate variability survey (\citealt{2015A&A...573A.103S, 2016A&A...588A.128F}), and consist of a large number of observations at varying hour angles, dates, and atmospheric conditions for each cluster.

Star positions were extracted using the \texttt{DAOPHOT} program \citep{1987PASP...99..191S}, using the best 1\% of exposures. Reference stellar astrometry was taken from a number of sources, listed in Table~\ref{tab:astrometricfits}. To account for pointing offsets between different observations, we cross-correlated each image with a reference image of the same cluster that was taken in good conditions and was well centred on the cluster.

We transformed the reference star positions to the measured positions using a four-parameter fit: the detector plate scale, in mas/px; the detector rotation eastwards from North; and the $(x, y)$ coordinates of the cluster centre for each image. For each iteration in the fitting process, the reference star positions were transformed and each measured star was paired with the nearest reference star. The data were fitted using the least-squares method, with the distances between each pair of stars being minimized. We assumed that the effect of proper motions was negligible, and that any effect would average out due to the large number of reference stars. 

A correction was made for atmospheric dispersion in the TCI data, which modifies the apparent position of stars,  following Eq.~12 of \citet{1998PASP..110..738G} and utilising their \texttt{slaRefco} routine to determine the required coefficients. Dispersion varies slightly with meteorological conditions, but archival meteorological data for the La Silla observatory during much of 2015 is missing\footnote{No data is returned from the archive at \url{http://archive.eso.org/wdb/wdb/asm/ambient_lasilla/form}}. We therefore assumed typical atmospheric conditions of $T=10$ C, $P=770$ mbar, and a relative humidity of $10\%$. We assumed a temperature lapse rate of $0.0065$K m$^{-1}$ \citep{1998PASP..110..738G}, an observatory altitude of $2340$ m and latitude of $-29.264\degrees$, and a wavelength of $800$ nm for the observed light.

\begin{figure} 
	\includegraphics[width=\columnwidth,angle=0]{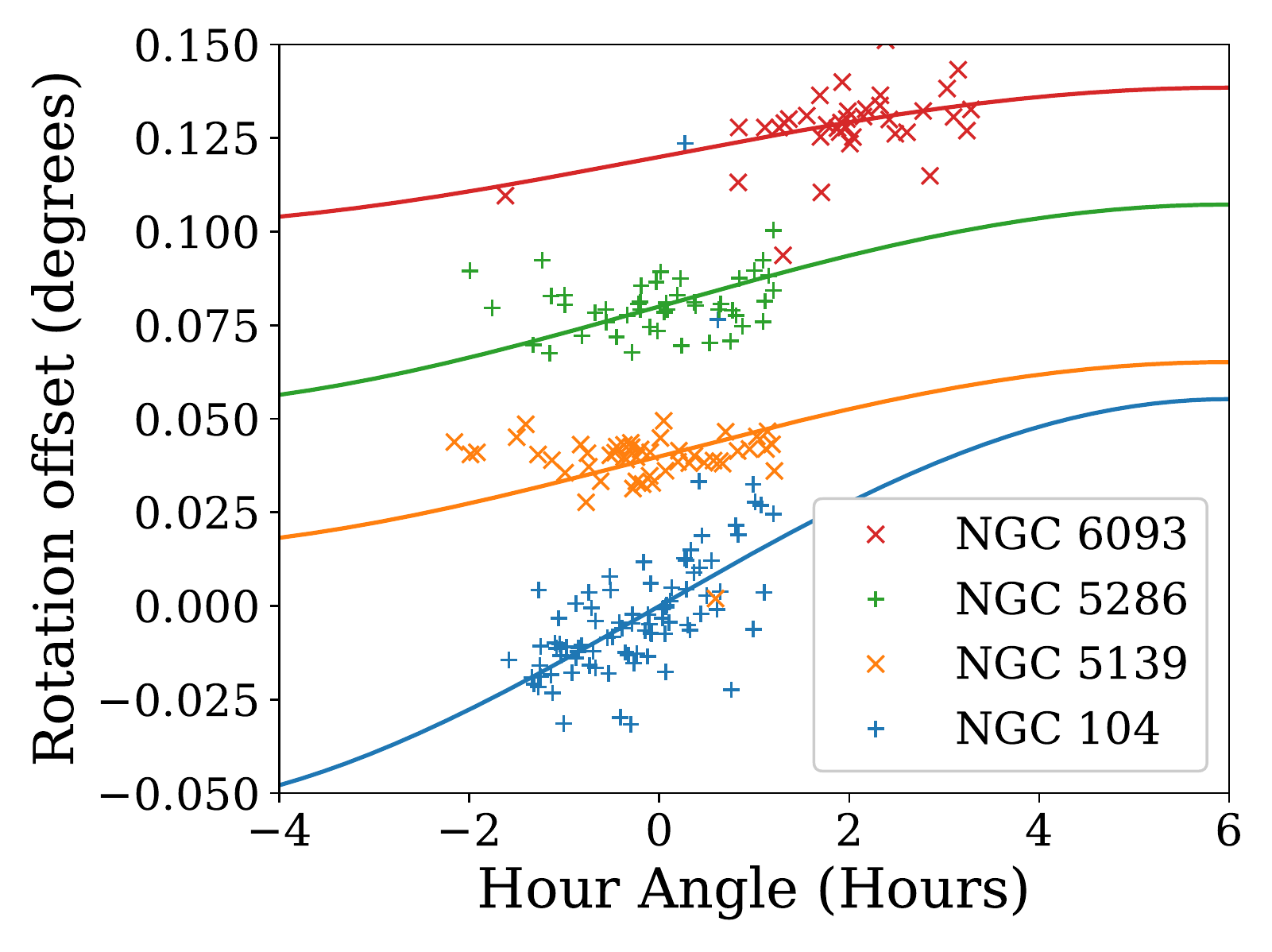}
	\caption{\label{fig:polaroffset} Variation of the derived detector orientation with the hour angle at time of observation. Coloured points show individual measurements for four globular clusters, with clusters being offset by $0.04\degrees$ for clarity. The solid sinusoidal lines show the predicted variation due to a polar axis offset of $0.017\degrees$. The change in detector orientation due to this effect increases towards the poles, with NGC 104 ($\delta=-72\degrees$) showing the highest amplitude. } 
\end{figure}

We inspected our astrometric results for correlations with a number of variables, and noted a variation with hour angle of up to $0.05\degrees$, shown in Fig.~\ref{fig:polaroffset}. The Danish 1.54m telescope is equatorially mounted, and so should not suffer from field rotation. However, if the polar axis is misaligned by some small angle $\theta$, this will result in field rotation that is most extreme towards the poles, and will vary sinusoidally with hour angle; both of these effects are seen. This effect can be modelled as the field rotation seen by an alt-az telescope situated at a latitude of $90-\theta$, and we fitted a polar axis offset of $\theta=0.017\pm0.003\degrees$. We corrected the orientation for each observation based on the telescope coordinates stored in the image FITS headers, and the final derived astrometric solutions for each cluster are given in Table~\ref{tab:astrometricfits}.

\begin{table*}
	\caption{\label{tab:astrometricfits} The results of astrometric fits to globular clusters. The `Ref.' column lists the source of the reference astrometric positions.}
	\centering
	\begin{tabular}{l c c c c c} \hline \hline 
		& \multicolumn{2}{c}{Red camera} & \multicolumn{2}{c}{Visual camera} \\
		Cluster & Scale (mas/px) & Rotation ($\degrees$) & Scale (mas/px) & Rotation ($\degrees$) & Reference \\
		\hline
		NGC 104 (47 Tuc)         & $89.03\pm0.07$ & $+1.222\pm0.012$ & $89.15\pm0.07$ & $-1.329\pm0.013$ & 1 \\ 
		NGC 104 (47 Tuc)         & $88.92\pm0.07$ & $+1.215\pm0.010$ & $89.04\pm0.07$ & $-1.335\pm0.013$ & 2 \\ 
		NGC 5139 ($\omega$ Cen)  & $88.87\pm0.05$ & $+1.206\pm0.004$ & $89.01\pm0.05$ & $-1.352\pm0.005$ & 2 \\ 
		NGC 5139 ($\omega$ Cen)  & $88.88\pm0.05$ & $+1.296\pm0.006$ & $89.02\pm0.05$ & $-1.258\pm0.007$ & 3 \\ 
		NGC 5139 ($\omega$ Cen)  & $88.87\pm0.05$ & $+1.200\pm0.005$ & $89.01\pm0.05$ & $-1.353\pm0.008$ & 4 \\ 
		NGC 5139 ($\omega$ Cen)  & $88.87\pm0.05$ & $+1.205\pm0.004$ & $89.01\pm0.05$ & $-1.352\pm0.006$ & 5 \\ 
		NGC 5286                 & $88.91\pm0.06$ & $+1.322\pm0.007$ & $89.06\pm0.06$ & $-1.236\pm0.008$ & 2 \\ 
		NGC 6093 (M80)           & $88.90\pm0.08$ & $+1.333\pm0.007$ & $89.03\pm0.08$ & $-1.222\pm0.006$ & 2 \\ 
		NGC 6121 (M4)            & $88.86\pm0.05$ & $+1.337\pm0.005$ & $89.01\pm0.05$ & $-1.216\pm0.012$ & 2 \\ 
		NGC 6121 (M4)            & $88.85\pm0.04$ & $+1.340\pm0.006$ & $89.00\pm0.05$ & $-1.213\pm0.011$ & 6 \\ 
		NGC 6388                 & $88.91\pm0.04$ & $+1.378\pm0.011$ & $89.13\pm0.15$ & $-1.177\pm0.027$ & 2 \\ 
		NGC 6541                 & $88.92\pm0.06$ & $+1.374\pm0.022$ & $89.10\pm0.09$ & $-1.182\pm0.011$ & 2 \\ 
		NGC 6652                 & $88.98\pm0.09$ & $+1.327\pm0.021$ & $89.15\pm0.09$ & $-1.213\pm0.040$ & 2 \\ 
		NGC 6656 (M22)           & $88.94\pm0.06$ & $+1.330\pm0.009$ & $89.10\pm0.10$ & $-1.232\pm0.027$ & 2 \\ 
		\hline
	\end{tabular}
	\tablefoot{The data from \citet{2008AJ....135.2055A} are available at http://www.astro.ufl.edu/\textasciitilde ata/public\_hstgc/}
	\tablebib{(1) \citet{2006ApJS..166..249M}; (2) \citet{2008AJ....135.2055A}; (3) \citet{2009A+A...507.1393B}; (4) \citet{2010AJ....140..631B}; (5) \citet{2010ApJ...710.1032A}; (6) \citet{2014A+A...563A..80L}.}
\end{table*}

\begin{figure} 
	\includegraphics[width=\columnwidth,angle=0]{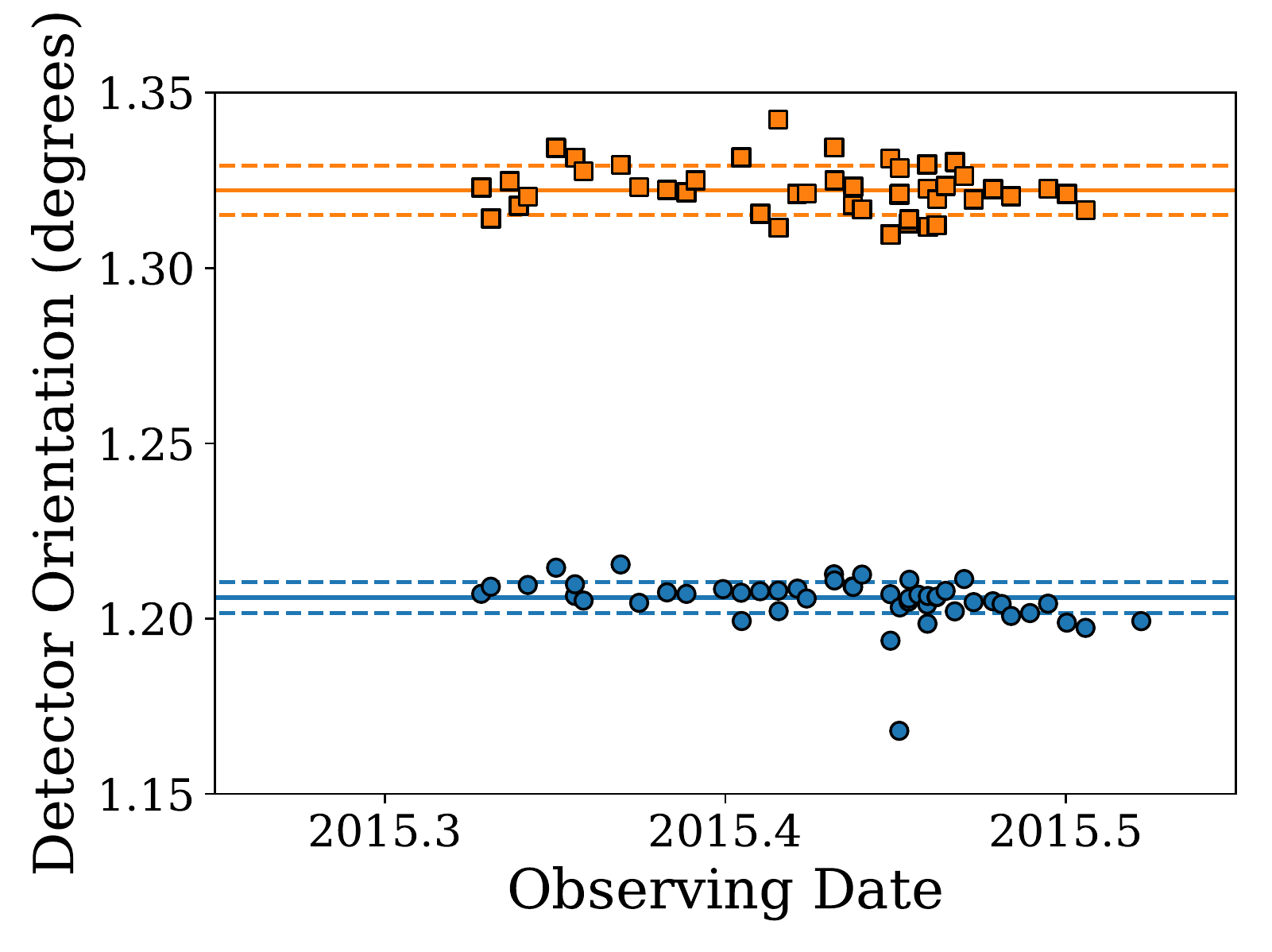}
	\caption{\label{fig:5139vs5286} The calculated detector orientation (measured in degrees eastwards of North) for the red camera, for all 2015 observations of NGC 5139 (blue circles) and NGC 5286 (orange squares). The mean and standard deviation of each set of observations are shown with solid and dashed lines respectively. The derived detector orientations for the two clusters are significantly different, far exceeding the scatter caused by uncertainties in the astrometric fitting process. }
\end{figure}

For both cameras, the derived detector scales from the individual clusters show good agreement. However, systematic offsets on the order of $0.1\degrees$ were found in the calculated detector orientation between different clusters, significantly larger than the RMS scatter of $\sim0.01\degrees$ for observations of the same cluster -- the offset between the individual measurements for NGC 5139 and NGC 5286 are clearly visible in Fig.~\ref{fig:5139vs5286}. The offsets in orientation are common between the cameras, with the relative angle between the two detectors being constant at $2.54\pm0.02\degrees$. 

In some cases, multiple sets of high quality astrometry were available for the same cluster, which should result in the same astrometric solution if the reference data have been correctly calibrated. For NGC 6121, the data in \citet{2014A+A...563A..80L} are matched to the UCAC4 reference frame, but agree well with the ACS Survey data. The two sets of data for NGC 104 use independently determined reference frames based on the HST guide stars, but also agree on detector orientation -- however, the detector scales are significantly different, and the scale we derive from \citet{2006ApJS..166..249M} is noticeably larger than for any other data set. For NGC 5139, \citet{2010AJ....140..631B} and \citet{2010ApJ...710.1032A} adopted the HST-based reference frame of \citet{2008AJ....135.2055A}, and the results from all three sets of data reflect this common reference frame; the positions in \citet{2009A+A...507.1393B} are ultimately linked back to the UCAC2 reference frame \citep{2009A&A...493..959B}, and are apparently orientated by $\sim0.1\degrees$ relative to the ACS Survey reference frame.

Rotation offsets between datasets have been noted previously for HST-derived data, with \citet{2010ApJ...710.1032A} noting a difference of $0.1\degrees$ between the ACS Survey and 2MASS for NGC 5139. The HST has been shown to match commanded roll angles to far better than $0.1\degrees$ \citep{2007acs..rept....7V}, and if the orientation differences do result from the reference data, we consider the most likely source of errors to be the astrometric calibration itself. For most of our sets of reference data, astrometry has been placed on an absolute reference frame using the HST guide stars \citep{2008AJ....135.2055A}. In normal operation, two guide stars are used by HST, with the astrometric uncertainties dominated by the accuracy of the Guide Star Catalog II (GSCII), which has an RMS scatter of $\sim0.3\arcsec$\ in stellar positions \citep{2006hstc.conf..417K}. We created a simple model of FGS star position measurements, based on the FGS parameters given in \citet{2016fgsi.book.....N} and the assumption that pairs of GSCII stars would be uniformly distributed across the available FGS detector space, with normally-distributed positional errors of $\sigma=0.3\arcsec$ on each star. From this model, we find an r.m.s. error of $0.025\degrees$ in the measured orientation of HST derived from two guide stars, with the range of offsets reaching up to $0.1\degrees$.

We conclude that the systematic offsets we see are due to the limited accuracy of our HST reference data, and that these are the limiting factor in determining our detector orientation and scale. Our final astrometric results are therefore a scale of $88.91\pm0.05$mas/px and orientation of $1.30\pm0.06\degrees$ eastwards of North for the red camera, and $89.07\pm0.05$mas/px and $-1.25\pm0.06\degrees$ eastwards of North for the visual camera. The fit results from \citet{2010AJ....140..631B} and \citet{2010ApJ...710.1032A} were not used when calculating these final values, as they share the same astrometric calibration as the ACS Survey, and therefore are not independent estimates of our detector orientation.

\subsection{Dichroic leak at 410nm} \label{sec:dichroic}

Light entering the TCI is split between the red and visual arms by a dichroic mirror with generally simple transmission/reflection properties, transmitting light redwards of 655nm to the red camera, and reflecting bluer light towards the visual camera. The transmission properties are less well behaved between 350 and 430nm, with the dichroic transmitting some light towards the red camera, the main transmission feature being located at 400nm. It was previously assumed that the only effect of the leak was a small contamination of any red camera photometry, which was included in the colour calibrations presented in Paper I.

\begin{figure} 
	\includegraphics[width=\columnwidth,angle=0]{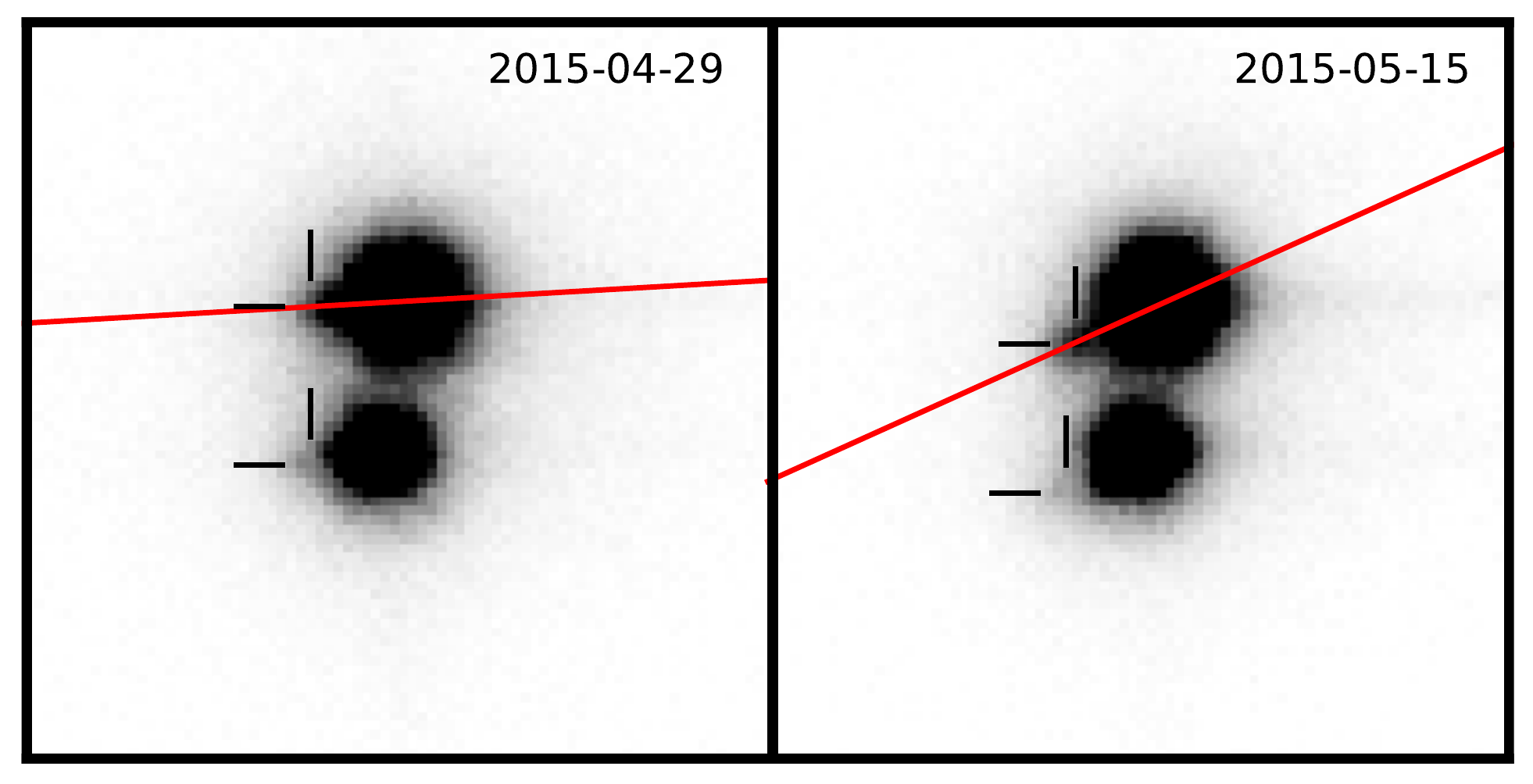}
	\caption{\label{fig:WASP85ghosts} Observations of the visual binary WASP-85 on 2015-04-29 (left) and 2015-05-15 (right). The data show `companions' associated with each of the stellar components, which are in fact secondary images of the two stars caused by the `blue leak' in the TCI dichroic. The separations of the secondary images increase with airmass. The position angles vary with the parallactic angle, which is indicated by the red lines. } 
\end{figure}

As the TCI does not have an atmospheric dispersion corrector (ADC), the apparent position of a star will change with wavelength. Given a wavelength difference of approximately 400nm between the blue leak and the bulk of the light reaching the red camera, it is possible for a star to appear twice on an image. At a zenith distance of $40\degrees$ (airmass 1.3) and typical atmospheric conditions,\footnote{Conditions for La Silla observatory assumed to be a temperature of 10C, atmospheric pressure at 770 hPa, 20\% humidity, and a temperature lapse rate of 0.0065K/m \citep{1998PASP..110..738G}} the secondary image formed by the blue light will be separated from the main image of the star by an arcsecond, and hence is easily resolved with the TCI. The secondary image will appear at a position angle corresponding to the parallactic angle $q$ (when measured from main image to secondary image, $q+180$\degrees). Fig.~\ref{fig:WASP85ghosts} shows two observations of the visual binary WASP-85, with both components having visible secondary images that vary in position with the observing conditions. Atmospheric dispersion also causes the main PSF of the target to be elongated, reducing our sensitivity to companions that fall along the parallactic angle.

\begin{table*}[htbp]
	\caption{\label{tab:comps2sigma} Physical properties of companion stars with distances consistent with physical association to planet host stars.}
	\centering
	\begin{tabular}{l c r@{\,$\pm$\,}l r@{\,$\pm$\,}l r@{\,$\pm$\,}l c l} \hline \hline 
		System & Comp. ID & \multicolumn{2}{c}{Proj. Sep.\tablefootmark{a} (au)} & \multicolumn{2}{c}{\drtci} & \multicolumn{2}{c}{Temperature (K)} & Evidence\tablefootmark{b} & References \\
		\hline
		CoRoT-2  & 1 & 920   & 90   & 2.85 & 0.04 & 3660 & 90  & S & 1, 2, 3, 4, 5, 6, 7 \\
		CoRoT-22 & 8 & 5400  & 500  & 3.88 & 0.03 & 3590 & 90  & - & - \\
		CoRoT-24 & 1 & 5400  & 600  & 4.78 & 0.08 & 3500 & 300 & - & - \\
		CoRoT-24 & 3 & 5700  & 600  & 5.12 & 0.10 & 3700 & 300 & - & - \\
		CoRoT-28 & 37 & 13200 & 1300 & 1.80 & 0.03 & 4030 & 140 & - & - \\
		HAT-P-30 & 1 & 750   & 80   & 4.34 & 0.04 & 3800 & 130 & P & 1, 7, 8, 9, 10, 11 \\
		HAT-P-35 & 1 & 500   & 100  & 4.09 & 0.11 & 3800 & 500 & P & 1, 7, 12 \\
		HAT-P-35 & 2 & 4800  & 500  & 4.72 & 0.04 & 3560 & 700 & P & 1, 7 \\
		HAT-P-41 & 1 & 1280  & 130  & 3.39 & 0.04 & 4370 & 120 & C & 1, 5, 6, 12 \\
		HATS-1   & 1 & 3500  & 400  & 8.09 & 0.06 & 3130 & 100 & - & 1 \\
		HATS-10  & 2 & 5800  & 600  & 7.87 & 0.06 & 3090 & 150 & - & - \\
		HATS-12  & 13 & 19000 & 1900 & 6.26 & 0.06 & 3440 & 100 & - & - \\
		HATS-12  & 21 & 25000 & 2500 & 7.68 & 0.10 & 3400 & 200 & - & - \\
		HATS-14  & 1 & 3900  & 400  & 7.13 & 0.11 & 3300 & 300 & - & - \\
		K2-02    & 1 & 490   & 50   & 7.45 & 0.13 & 7300 & 200 & P & 7, 13 \\
		K2-27    & 1 & 710   & 70   & 5.38 & 0.06 & 3150 & 80  & - & 7, 14 \\
		K2-38    & 1 & 2060  & 210  & 9.02 & 0.17 & 3400 & 500 & - & - \\ 
		KELT-15  & 1 & 1940  & 200  & 6.03 & 0.05 & 3200 & 70  & - & - \\
		WASP-8   & 1 & 380   & 40   & 3.61 & 0.04 & 3550 & 80  & P & 1, 7, 11, 15 \\
		WASP-19  & 10 & 4600  & 500  & 8.02 & 0.10 & 3260 & 180 & - & 1 \\ 
		WASP-19  & 11 & 4700  & 500  & 6.54 & 0.06 & 3460 & 80  & - & 1 \\ 
		WASP-24  & 1 & 5400  & 600  & 8.19 & 0.09 & 3444 & 226 & - & 1 \\
		WASP-26  & 1 & 3900  & 400  & 2.65 & 0.04 & 4390 & 160 & P & 1, 16 \\
		WASP-36  & 1 & 2000  & 200  & 4.55 & 0.04 & 3490 & 60  & - & 1, 6, 12, 17 \\
		WASP-36  & 3 & 5300  & 500  & 5.43 & 0.05 & 3460 & 60  & - & 1, 17 \\
		WASP-45  & 1 & 920   & 100  & 6.33 & 0.06 & 3020 & 90  & P & 1 \\
		WASP-49  & 1 & 450   & 50   & 5.08 & 0.07 & 3220 & 100 & - & 1, 18 \\
		WASP-54  & 1 & 1580  & 160  & 7.94 & 0.08 & 3050 & 120 & - & 1 \\
		WASP-55  & 1 & 1330  & 130  & 5.32 & 0.04 & 3280 & 80  & - & 1 \\
		WASP-66  & 1 & 6800  & 700  & 8.81 & 0.13 & 3330 & 150 & - & 1 \\
		WASP-68  & 2 & 5000  & 500  & 8.20 & 0.09 & 3020 & 80  & - & - \\
		WASP-70  & 1 & 700   & 70   & 2.12 & 0.04 & 4900 & 300 & CPS & 1, 19 \\
		WASP-77  & 1 & 330   & 30   & 1.55 & 0.03 & 4600 & 200 & CPS & 1, 7, 20 \\
		WASP-85  & 1 & 175   & 17   & 0.99 & 0.08 & 5200 & 300 & PS & 6, 7, 21, 22, 23 \\
		WASP-87  & 4 & 2600  & 300  & 1.95 & 0.04 & 5300 & 300 & C & 24 \\
		WASP-94  & 1 & 3400  & 300  & 0.36 & 0.04 & 6400 & 500 & PS & 7, 25, 26 \\
		WASP-104 & 1 & 1320  & 130  & 6.07 & 0.05 & 3010 & 80  & CP & - \\
		WASP-106 & 2 & 6200  & 600  & 8.09 & 0.12 & 3400 & 400 & - & - \\
		WASP-108 & 4 & 2200  & 200  & 7.49 & 0.11 & 3150 & 150 & - & - \\
		WASP-109 & 6 & 7300  & 700  & 8.01 & 0.09 & 3010 & 60  & - & - \\
		WASP-111 & 1 & 1400  & 140  & 4.43 & 0.04 & 3710 & 60  & P & - \\
		WASP-121 & 1 & 2000  & 200  & 7.49 & 0.08 & 3060 & 110 & - & - \\
		WASP-123 & 1 & 1030  & 100  & 5.15 & 0.06 & 3200 & 60  & - & - \\
		WASP-129 & 2 & 1460  & 150  & 5.65 & 0.05 & 3300 & 70  & - & - \\
		WASP-133 & 4 & 9000  & 900  & 8.59 & 0.16 & 3500 & 500 & - & - \\
		\hline	\end{tabular}
	\tablefoot{Companions with distances consistent at $2\sigma$ or better are included, except in the case of significant evidence against physical association (e.g. discrepant proper motions). No companions were found to have discrepant distance measurements from literature photometry or spectroscopy. We also exclude most companions to CoRoT-04, CoRoT-22, HAT-P-45, and HATS-11 from this table, due to a high fraction of background stars being identified with consistent distances. \\
		\tablefoottext{a}{Projected Separation.}
		\tablefoottext{b}{Codes in this column indicate the further evidence for physical association: C -- Colour-distance comparisons using non-TCI photometry; P -- Common proper motion; S -- Spectroscopic characterisation.}}
	\tablebib{(1) Paper I \citep{HITEP1}; (2) \citet{2008A&A...482L..21A}; (3) \citet{2011A&A...532A...3S}; (4) \citet{2013MNRAS.433.2097F}; (5) \citet{2015A&A...575A..23W}; (6) \citet{2015A&A...579A.129W}; (7) Washington Double Star catalog \citep{2001AJ....122.3466M}; (8) \citet{2011AJ....142...86E}; (9) \citet{2013AJ....146....9A}; (10) \citet{2013A&A...559A..71G}; (11) \citet{2015ApJ...800..138N}; (12) \citet{2016ApJ...827....8N}; (13) \citet{2015ApJ...800...59V}; (14) \citet{2015ApJ...809...25M}; (15) \citet{2010A&A...517L...1Q}; (16) \citet{2010A&A...520A..56S}; (17) \citet{2012AJ....143...81S}; (18) \citet{2016A&A...587A..67L}; (19) \citet{2014MNRAS.445.1114A}; (20) \citet{2013PASP..125...48M}; (21) \citet{2015EPSC...10..603B}; (22) \citet{2016AJ....151..153T}; (23) \citet{2016AJ....151..159S}; (24) \citet{2014arXiv1410.3449A}; (25) \citet{2014A&A...572A..49N}; (26) \citet{2016ApJ...819...19T}.}
\end{table*}

\section{Lucky Imaging Results} \label{sec:results}

All detected companions are listed in Tables~\ref{tab:data15}~and~\ref{tab:data16}, covering our 2015 and 2016 observations respectively. Where targets were observed in 2014, we retain the original companion numbering system for ease of comparison. Newly detected companions are assigned numbers in order of increasing separation.

In the following short paragraphs, we consider in detail companions with colours and magnitudes indicating that they are likely to be bound to the target host stars, using archival photometry and astrometry where possible. Also mentioned are companions that are particularly bright or close to the planet host stars, due to the increased fraction of contaminating light that these stars produce. In Table~\ref{tab:comps2sigma}, we provide a summary for companions whose distances are consistent to within $2\sigma$ of the relevant planet host star, excluding companions where there is other available evidence (e.g. discrepant proper motions). Due to stellar crowding and high reddening resulting in many stars having apparently consistent distances, this table does not include the majority of companions to CoRoT-04, CoRoT-22, and HAT-P-45. 

In considering individual companions of interest, archival data were sourced from a variety of catalogues. Where available, we used ground-based proper motions from the NOMAD \citep{2004AAS...205.4815Z}, PPMXL \citep{2010AJ....139.2440R}, UCAC4 \citep{2013AJ....145...44Z}, and URAT1 \citep{2015AJ....150..101Z} catalogues to assess whether pairs of stars showed common proper motion, but we note that many of our detected companions are too faint to have reliable proper motions, if any at all. We also used archival position measurements to compare to our own, with 2MASS \citep{2003yCat.2246....0C} positions dating from around epoch 2000, and Gaia Data Release 1 (Gaia DR1, \citealt{2016A&A...595A...1G}) positions at epoch 2015.0.

The 2MASS catalogue also includes $JHK$ photometry for some companions, although we were rarely able to derive properties for companions from 2MASS alone, as dwarf stars below $\sim0.7\Msun$ show little variation in their $JHK$ colours, and the faintness of most detected companions resulted in large photometric uncertainties. More rarely, resolved photometry was available in the optical, mainly from the SDSS DR9 \citep{2012ApJS..203...21A} and APASS \citep{2016yCat.2336....0H} catalogues. Finally, some of our targets are included in the Washington Double Star catalog (WDS, \citealt{2001AJ....122.3466M}), which includes historical astrometry and photometry, with records extending back to the 19th century in a few cases.

{\bf CoRoT-2.}
Companion 1 is a bright star near CoRoT-2 that was noted by \citet{2008A&A...482L..21A}, with several lucky imaging observations since (\citealt{2013MNRAS.433.2097F, 2015A&A...575A..23W, 2015A&A...579A.129W}, and Paper I). Additionally, \citet{2011A&A...532A...3S} characterised the companion spectroscopically, measuring a temperature between 3900K and 4100K (95\% confidence interval), and finding its radial velocity to be consistent with CoRoT-2. We measure a somewhat lower temperature of $3660\pm90$K for the companion, and find no other companions consistent with being physically associated.

{\bf CoRoT-22.}
Of the companions observed in 2016, it is unsurprising that a significant portion have distances that are consistent with CoRoT-22, given its relatively large distance of nearly 600pc \citep{2014MNRAS.444.2783M} and location in the galactic plane. We found no companions closer than a star located $3.3\arcsec$\ to the North, which was previously imaged by \citet{2014MNRAS.444.2783M} using VLT/NACO. One notable companion, number 8, is located $8.4\arcsec$\ to the Southwest, $3.9$ mag fainter in $\rtci$, with an estimated distance that differs by only $0.2\sigma$ from that of CoRoT-22, perhaps indicating that it is a bound companion in a wide orbit. However, due to the high probability of chance alignments, further evidence is required before any definite conclusions can be made.

{\bf CoRoT-28.}
From our 2015 observations, we detected a number of faint background stars within the CoRoT photometric mask that were not identified by \citet{2015A&A...579A..36C}. The brightest companion falling within the photometric mask, companion 3, is $5.9$ mag fainter in $\rtci$ at a separation of $5.9\arcsec$, and all other stars are at least $6.4$ magnitudes fainter in the same filter. Companion 37 is the brightest star detected in our observations, at a separation of $22.5\arcsec$, is the only companion with a colour consistent with being bound; however, neither the NOMAD, PPMXL, or UCAC4 catalogues support common proper motion, and so this star is likely unassociated with CoRoT-28.


{\bf HAT-P-27.}
A close companion located $0.65\arcsec$\ from HAT-P-27 was announced by \citet{2015A&A...579A.129W} and also observed by \citet{2016ApJ...827....8N}. However, our observations do not sufficiently resolve these two stars due to the companion's faintness, with the companion being visible as a small distortion in the 2015 data, and obscured by the PSF elongation caused by atmospheric dispersion in the 2016 data.

{\bf HAT-P-30.}
Both companion stars observed in 2014 were reobserved in 2015 and 2016. Companion 1 has previously been analysed by \citet{2011AJ....142...86E}, \citet{2013AJ....146....9A} \citet{2013A&A...559A..71G}, and \citet{2015ApJ...800..138N}. Common proper motion was preferred by the analysis in Paper I and that of \citet{2015ApJ...800..138N}, and our new separation measurements are consistent with this conclusion. We derive a colour temperature of $3800\pm130$K, and the measured flux ratios in $\rtci$ and $\vtci$ correspond to those expected of a bound companion, and so we conclude that these two stars are physically associated.

Companion 2 is included in the WDS as a `C' component in the system. In paper I, we noted that 2MASS $J$ and $H$ magnitudes for the companion made it inconsistent with being bound. Our two colour 2015 data are in agreement with this conclusion, with a colour temperature of $5700\pm500$K, the star being much too faint to be a bound companion at this temperature. In addition, the URAT1 catalogue indicates that the two stars do not exhibit common proper motion. 

{\bf HAT-P-35.}
All four companion stars found in Paper I were reobserved in 2015 and 2016. Companion 1 is the closest at $0.93\arcsec$, and has also been observed by \citet{2015A&A...579A.129W} and \citet{2016ApJ...827....8N}, with the latter work confirming common proper motion, but this companion is not sufficiently resolved in either of our datasets from 2015 or 2016 for reliable astrometry. Two colour photometry indicates a temperature in the range $3700-4200$K, somewhat higher than the values of $3525\pm76$K and $3563\pm70$K derived by \citet{2016ApJ...827....8N}.

For companion 2 at $9\arcsec$, we derive a temperature of $3560\pm70$K for this star, with its measured flux ratios being consistent with a bound companion to HAT-P-35. Resolved SDSS and 2MASS photometry are available for this companion, with the BT-Settl model isochrones \citep{2012RSPTA.370.2765A}, indicating a $M=0.5\Msun$ star ($\Teff=3680$K) at the same distance as HAT-P-35, using the \citet{2012AJ....144...19B} age and distance estimates for the system. Combining 2MASS and Gaia astrometry with our own, we find no significant trends in separation or position angle, though we note that HAT-P-35's proper motion is low. The NOMAD catalogue gives proper motions for the star that are consistent at about $1\sigma$, whilst the URAT1 catalogue clearly indicates common proper motion. It therefore seems likely that this companion is also physically associated with HAT-P-35, making the system a relatively unusual example of a hierarchical triple with three resolved components.

{\bf HAT-P-41.}
In Paper I, we reported the detection of a new companion at $1.0\arcsec$\ with a PA of $190\degrees$, the magnitude difference being measured as $\drtci=4.4$. However, despite the good seeing in our 2015 and 2016 observations, we do not re-detect this companion; similarly, the companion was not detected in \citet{2015A&A...575A..23W} or \citet{2015A&A...579A.129W}, despite our reported magnitude being above their $5\sigma$ detection limit.

Following the discovery that the TCI can produce secondary images of the target star, discussed in Sect.~\ref{sec:dichroic}, we calculated the expected properties of a secondary image for our 2014 observations\footnote{Weather conditions at time of observing were retrieved via ESO's La Silla Meteorology Query Form, available at \url{http://archive.eso.org/wdb/wdb/asm/meteo_lasilla/form}}. The predicted separation of $0.8\arcsec$\ and PA of $195\degrees$ correspond closely with our 2014 detection, and we therefore conclude that this was a spurious detection.

Companion 1 at $3.6\arcsec$ was previously observed by \citet{2015A&A...575A..23W}, \citet{2015A&A...579A.129W}, and \citet{2016ApJ...827....8N}, with the latter finding the star to be consistent with a bound companion based on its colours, although their proper motion analysis was inconclusive. With a temperature of $4470\pm120$ K, we find the companion's distance is consistent with HAT-P-41 at the $1\sigma$ level.

{\bf HAT-P-45.}
Due to its position near the galactic plane, a large number of faint, red stars were observed near HAT-P-45 in 2015, many of which have consistent distances. It is highly likely that many, if not all of these stars are unassociated with HAT-P-45, and further evidence would be required to determine whether any are likely to be bound.

{\bf HATS-1.}
In 2015 we re-observed our previously detected companion 1, located at a separation of $11.5\arcsec$\ and $8$mag fainter in $\rtci$. The companion has a position measured in Gaia DR1, with no significant trend in separation or position angle being seen. With a colour temperature of $3130\pm100$K, the companion appears slightly fainter than expected of a star at the same distance as HATS-1 at the $2\sigma$ level.

We also detected a new companion located $6\arcsec$ to the South of HATS-1, and $10$mag fainter in $\rtci$. This star falls almost directly beneath one of the diffraction spikes originating from HATS-1, and as a result our photometry is badly contaminated; whilst we do measure a very blue colour for this companion ($\vtci-\rtci=-0.5\pm0.3$), we consider this unreliable.

Both of these companions are separate from the very wide common proper motion companion identified by \citet{2014MNRAS.439.1063M}.

{\bf HATS-2.}
The close companion at $1.1\arcsec$\ found in 2014 was re-observed in 2015 and 2016. The UCAC4, NOMAD, and PPMXL catalogues all agree that HATS-2 is moving predominantly in RA at a rate of approximately $-45$mas/yr, which would lead to increasing separation and position angle given the current position of the companion. Our astrometric measurements do indicate positive trends, but these are not statistically significant ($0.3\sigma$ in separation, $0.6\sigma$ in position angle). Analysis of the colour of the companion indicates that it is much too faint to be physically associated, its flux ratios being $4\sigma$ too faint to match our measured temperatures of $4300\pm300$K (2015 observation) and $3900\pm220$K (2016).

{\bf HATS-3.}
We reobserved the close companion at $3.6\arcsec$\ in 2016, deriving a temperature of $4000\pm300$K, with the star therefore being $4\sigma$ too faint to be physically associated.

{\bf HATS-11.}
Located in a fairly crowded field, we detected 28 companion stars in an observation in 2016. Companion 1 is located at only $1.4\arcsec$, and is $4.6$ mag fainter in $\rtci$. This star is too faint at the $2\sigma$ level, with a temperature of $3920\pm160$K.

Companions 2, 3, and 7 have colours and flux ratios that give distances that are consistent with HATS-11 to within $1\sigma$. Given HATS-11's distance of $900$pc and a location near the galactic plane, the probability of chance alignments of stars with calculated distances that agree within our uncertainties is relatively high.

{\bf HATS-14.}
Of the five stars detected in 2016, only companion 1 at $7.8\arcsec$\ has a colour and magnitude consistent with being a bound object at less than $2\sigma$. Companion 3 has a diffuse, extended profile, and is likely a background galaxy.

{\bf HATS-26.}
Two objects were detected in our 2016 observations. Companion 1 has a thin, extended profile, which we identify as an edge-on spiral galaxy. When treated by our reduction pipeline as a star, the centre of this galaxy is measured to be $\sim5\arcsec$ from HATS-26 and $7$ mag fainter, with a relatively blue colour corresponding to a black-body temperature of $6700$K. Visual inspection shows that the galaxy's orientation and extended profile result in emission as close as $2\arcsec$ to HATS-26, and with the planet being a tempting target for transmission spectroscopy \citep{2016AJ....152..108E}, it is worth noting the presence of such a close background object.

{\bf HATS-27.}
Companion 5 has similar brightness to HATS-27 at a separation of $21.9\arcsec$, with colours marginally consistent with being a bound object, the companion appearing to be nearer than HATS-27 at $2\sigma$. Being well separated and of similar brightness to HATS-27, the star is resolved in various catalogues. The companion's colours in the APASS9 and 2MASS catalogues indicate a $\mathrm{K}3/4$ star, with HATS-27 being $\mathrm{F}5/6$; the expected difference in brightness would be on the order of $3$ mag in $V$, much larger than the APASS9 difference of $0.90\pm0.06$ mag.

The NOMAD, PPMXL, and UCAC4 catalogues disagree significantly with each other regarding the proper motions of the two stars, but none indicate that the pair show common proper motion. With no evidence favouring the bound scenario, we conclude that this companion is an unassociated foreground star.

{\bf K2-02 (HIP 116454/EPIC 60021410).} 
In their analysis of this system, \citet{2015ApJ...800...59V} noted a faint companion star, with archival imaging clearly showing that the two stars have common proper motion. SDSS photometry indicated that, whilst $7$ magnitudes fainter, the companion is bluer than HIP 116454, leading \citet{2015ApJ...800...59V} to characterise the faint star as a white dwarf. Our two-colour imaging supports this hypothesis, with our derived colour temperature of $7300\pm200$K being in agreement with the SDSS-derived value of $7500\pm200$K.

{\bf K2-21 (EPIC 206011691).}
\citet{2015ApJ...811..102P} presented Keck NIRC2 observations in the $K_{\rm cont}$ filter as part of their characterisation of the K2-21 system, finding no companion stars with $\Delta K < 7.5$ at separations beyond $0.5\arcsec$. We detect a single faint companion at $2.99\pm0.07\arcsec$, $8.8\pm0.5$ mag fainter in $\rtci$ and undetected in $\vtci$. Whilst the sensitivity curve and image presented in \citet{2015ApJ...811..102P} cover only the inner two arcseconds, the observations did cover the position of our newly discovered companion, with no star being detectable at this position (D. Ciardi, priv. comm.); this non-detection in $K_{\rm cont}$ places limits on the companion's properties. Using the \citet{2015A&A...577A..42B} isochrones, K2-21's temperature of $4043\pm375$K \citep{2015ApJ...811..102P}, and assuming our $\rtci$ magnitude difference to be comparable to $I_c$, we estimate K2-21 to have $I_c-K=1.8$mag, and hence the companion $I_c-K<2.5$. This corresponds to a main sequence star with $\Teff\gtrapprox3000$K, which would appear much brighter if bound to K2-21; we therefore conclude that the companion is an unassociated background star.

{\bf K2-24 (EPIC 203771098).}
Similarly to K2-21, a Keck NIRC2 observation with no detected companion stars was presented in \citet{2016ApJ...818...36P}, using the Br-$\gamma$ filter. We detect several companions, including a faint star at $3.82\pm0.09$\arcsec. With $\Delta \rtci = 8.66\pm0.18$ and $\Delta \vtci = 8.88\pm0.16$, the companion has a relatively similar colour to K2-24; assuming the same brightness difference in Br-$\gamma$, the companion would have been near the detection limit of the observations of \citet{2016ApJ...818...36P}. With a colour temperature of $5400\pm600$K, the star is clearly much too faint to be a bound main sequence companion.

{\bf K2-27 (EPIC 201546283).}

\citet{2015ApJ...809...25M} detected a companion star to K2-27 in a PHARO observation taken on 2015 February 4. The companion was measured to be at a separation of $2.98\pm0.05\arcsec$\ with a position angle of $177.7\pm0.4\degrees$, $3.72\pm0.06$ mag fainter in $K_s$ (Montet 2017, priv. comm.). We reobserved the companion in 2016 and measured differential magnitudes of $\Delta \vtci=6.97\pm0.13$ mag and $\Delta \rtci=5.38\pm0.06$ mag, corresponding to a temperature of $3150\pm80$K and a photometric distance consistent with a bound companion to K2-27. Using the \citet{2015A&A...577A..42B} isochrones and our measured temperature, we find that the $K_s$ magnitude difference also supports this hypothesis.

Although not close enough to be included in our observations, we note the presence of a nearby bright star visible on DSS images, EPIC 201546361, at a separation of $32\arcsec$\ to the West. The PPMXL and UCAC4 catalogues show the two stars to be moving southwards, with their proper motions consistent well within $1\sigma$. The NOMAD catalogue gives a similar direction of motion, but with a larger $2\sigma$ discrepancy between the individual motions. The URAT1 catalogue also supports common proper motion, but disagrees significantly with all other catalogues as to the direction of motion, showing the pair of stars to be moving to the Southeast; it is not clear why these values are so discrepant.

As the latter pair of stars are both bright and are well-separated from each other, photometry is available in a number of literature surveys. To determine temperatures and distances for the two stars, we fitted the ATLAS9 model atmospheres \citep{2004astro.ph..5087C} to the literature photometry the two stars. We used the GALEX near-UV magnitude \citep{2012yCat.2312....0B}; APASS $B$, $V$, $r$, and $i$ magnitudes \citep{2016yCat.2336....0H}; 2MASS $J$, $H$, and $K_s$ magnitudes \citep{2006AJ....131.1163S}; and WISE $W1$ and $W2$ magnitudes \citep{2012yCat.2311....0C} in our fit. We fixed $\logg = 4.5$ for the fits, and zero reddening was assumed. The effect of the faint companion to K2-27 was neglected, the flux contribution being only $6\%$ in $K_s$ and less at shorter wavelengths.

For K2-27, we determined a temperature of $5250\pm130$K, consistent with the literature value of $5320\pm70$K \citep{2015ApJ...809...25M}, and a distance of $236\pm19$pc. For EPIC 201546361 we derive a temperature of $5880\pm140$K and distance of $220\pm15$pc. As the distances to the two stars are consistent with one another, as well as common proper motion being favoured by all catalogues we inspected, we conclude that it is likely that these two bright stars are physically associated. This hypothesis will be easily tested with future Gaia data releases, which will provide accurate parallaxes and proper motions for these two bright stars.

{\bf K2-31 (EPIC 204129699).}
We detected a faint star located $8.4\arcsec$\ from K2-31 in our 2016 observations, $7.8$ mag fainter in $\rtci$. The companion is included in both 2MASS and Gaia DR1, and there is a highly significant trend of increasing separation over time, leading us to conclude that this companion is likely to be a background star.

{\bf K2-44 (EPIC 201295312).}
A companion star located 8 arcsec. from K2-44 is included in numerous catalogues, and was observed by \citet{2016ApJS..226....7C} using NIRC2 on Keck and PHARO on the Palomar 200 inch on unspecified dates, with separations and relative magnitudes in $K$ being reported. We observed this companion in 2016, with its colours and flux ratios indicating that it is likely a foreground object. Combining our astrometry with measurements from 2MASS and Gaia DR1, there are significant trends of increasing separation ($8.3\sigma$) and position angle ($3.7\sigma$), indicating that the two stars are unassociated. This conclusion is further supported by the URAT1 catalogue, which gives proper motions for the two stars that are inconsistent at the $4\sigma$ level.

{\bf KELT-15.}
Of the nearby stars detected in 2016, only companion 1 is consistent with being bound to KELT-15. Located $6.13\arcsec$\ away at a position angle of $283\degrees$, we determine a temperature of $3210\pm70$K which is consistent with a bound object at $1\sigma$, the companion being somewhat brighter than expected of a bound companion.

{\bf WASP-8.}
The faint, close companion to WASP-8 was noted by \citet{2010A&A...517L...1Q}, with measurements of the pair being recorded in the WDS as far back as 1930. The companion was detected in Paper I, and also imaged by \citet{2015ApJ...800..138N}, with all observations indicating a bound companion. We derive a temperature of $3550\pm80$K for the companion, somewhat cooler than our Paper I temperature of $3740\pm100$K, but in good agreement with the $3380-3670$K range quoted by \citet{2015ApJ...800..138N}.

{\bf WASP-26.}
\citet{2010A&A...520A..56S} noted a bright visual companion to WASP-26 at a separation of $15\arcsec$, and using archival photometry found the star to have a distance consistent with WASP-26. Furthermore, no relative motion was found when comparing the positions of the two stars between the 1950s Palomar Observatory Sky Survey (POSS) and the 2MASS survey, indicating common proper motion. Our two-colour photometry supports the conclusion that the two stars have a common distance, with our temperature of $4510\pm210$K being in agreement with the \citet{2010A&A...520A..56S} determination of $4600\pm120$K. We find no evidence of a trend in separation or position angle, and also note that the PPMXL catalogue gives proper motions for the two stars that agree at the $2\sigma$ level.

{\bf WASP-36.}
Four companion stars to WASP-36 were noted by \citet{2012AJ....143...81S}, who concluded that none of the stars were physically associated based on proper motion catalogues. However, we find that the measurements in ground-based NOMAD, PPMXL, and UCAC4 catalogues suffer from significant confusion between the sources, with multiple spurious entries and no consistent proper motion measurements for any of the companions, and no proper motion determination at all for the closest companion. The 2MASS catalogue contains resolved $JHK$ measurements for the companions, which are all consistent with temperatures between $2000$K and $4000$K -- more accurate determination is not possible due to the large photometric uncertainties, and the small change in $JHK$ colours with temperature in this regime.

Two-colour imaging indicates that companion 4 is likely to be a foreground object, whilst companion 2 appears to be a background star; the remaining companions, 1 and 3, both have colours that are consistent with being physically associated with WASP-36 within $1\sigma$. Whilst companion 1 has also been observed by \citet{2015A&A...579A.129W} and \citet{2016ApJ...827....8N}, it is not yet possible to draw any conclusions from the common proper motion analysis.

{\bf WASP-45.}
We reobserved the previously detected companion star in 2016, and find that its two-colour photometry is consistent with a bound object, with a temperature of $3020\pm90$K. Astrometry for the pair exists in Gaia DR1, with WASP-45 also having a measured proper motion based on the Tycho-Gaia Astrometric Solution (TGAS), moving at $70$mas/yr at a position angle of $131\degrees$. If the companion were a stationary background star, a change of $0.14\arcsec$\ would be expected between our 2014 and 2016 observations -- this is not the case, with no significant trend present in either separation or position angle. We therefore conclude that these two stars form a physical binary.

{\bf WASP-49.}
A close companion star at $2.2\arcsec$\ was discovered by \citet{2016A&A...587A..67L}, and also detected in our 2014 observations. \citet{2016A&A...587A..67L} also obtained a spectrum of the companion relative to WASP-49, finding it to be redder than WASP-49, but not characterising it further. We derive a colour temperature of $3230\pm100$K for the companion, with the measured flux ratios being slightly inconsistent with a bound object at $1\sigma$, the companion being brighter than expected. Gaia DR1 resolves the two stars, including TGAS proper motions for WASP-49, but with relatively high uncertainties on the position of the companion. There is a slight decreasing trend in separation and increasing trend in position angle, as expected of a stationary star given WASP-49's motion, but these trends are not statistically significant; the situation for this companion is therefore inconclusive. From two-colour photometry, we conclude that all other companion stars, including that at $9\arcsec$\ reported by \citet{2012A&A...544A..72L}, are background objects.

{\bf WASP-54.}
We re-observed a companion star at $5.7\arcsec$\ in both 2015 and 2016. We determine a temperature of $3000\pm110$K, and find that it is consistent with being a bound object. No significant trend is seen in separation or proper motion, but more measurements are required to definitively confirm common proper motion.

{\bf WASP-55.}
Companion 1 at $4.3\arcsec$\ was re-observed in 2015 and 2016, with two-colour photometry from both years indicating a temperature of $3340\pm90$K. Our measured flux ratios are consistent with a physically associated companion at $1\sigma$, the companion being slightly brighter than expected. WASP-55's proper motion is too low to confirm common proper motion from the available data, but no significant trends are seen in our astrometric measurements.

{\bf WASP-70 and WASP-77.}
The WASP-70 and WASP-77 systems both have bright, nearby companions with similar properties to the planet host stars. As in Paper I, we support the conclusion that WASP-70AB and WASP-77AB are binary systems. From our two-colour photometry of both B components, we measure temperatures that are slightly lower than the literature values. For WASP-70B, we derive a temperature of $4510\pm230$K, compared to the previous measurement of $4900\pm200$K by \citet{2014MNRAS.445.1114A}; similarly, we measure $4570\pm240$K for WASP-77B, compared to a previous value of $4700\pm100$K \citep{2013PASP..125...48M}.

{\bf WASP-85.}
The stars in the WASP-85AB system were characterised by \citet{2015EPSC...10..603B}, measuring a spectroscopic temperature of $5250\pm90$K for the secondary star, in good agreement with the weighted mean of our 2015 and 2016 measurements of $5340\pm340$K. The companion has since been observed by \citet{2015A&A...579A.129W}, \citet{2016AJ....151..153T}, and \citet{2016AJ....151..159S}, although we note that the latter two papers are separate reductions of the same data. Combining these measurements with historical data recorded in the WDS, orbital motion is clearly seen, with a $15\degrees$ change in position angle since 1880, although no clear trend is visible in separation. The historic data are not sufficiently precise to allow us to place any meaningful constraints on the binary orbit, and further high-precision astrometry is required to fully determine the orbit.

{\bf WASP-87.}
\citet{2014arXiv1410.3449A} noted a bright star $8.2\arcsec$\ to the Southeast of WASP-87, and suggested that the two stars may form a bound WASP-87AB system, based on resolved 2MASS photometry indicating a temperature of $5700\pm150$K for the secondary. It was also stated that the two stars had ``similar'' proper motions in the UCAC4 catalogue, but we note that the differences are significantly larger than the quoted uncertainties. Additionally, there are significant differences between the proper motions quoted by the UCAC4, PPMXL (primary star only), and NOMAD catalogues, likely due to confusion between sources in a crowded field.

Based on our observations in 2015 and 2016, we find that WASP-87B appears slightly too faint compared to WASP-87A based on its colour temperature. However, our analysis does not account for the fact that WASP-87A is somewhat evolved, and hence appears brighter than predicted by our temperature-radius relationship -- we therefore conclude that these two stars are consistent with a bound pair.

{\bf WASP-94.}
WASP-94 is another system with a bright companion, with the addition of a non-transiting planet orbiting WASP-94B. Our measured temperature of $6400\pm500$K for the B component is in agreement with the \citet{2014A&A...572A..49N} determination of $6040\pm90$K. \citet{2016ApJ...819...19T} measured elemental abundances of the two stars, claiming that significant differences were seen, despite both stars having similar stellar properties and orbiting giant planets. Under the assumption that the differences were caused by planet formation, the authors hypothesised that additional stellar companions may have affected the planet formation process. Based on our sensitivity limits given in Table~\ref{tab:obslist16} and the temperatures of WASP-94 A and B (6198K and 6112K respectively, citealt{2016ApJ...819...19T}) we exclude bound companion stars hoter than 3800K beyond 0.8\arcsec, 3300K beyond 1.5\arcsec, and 2900K beyond 2.5\arcsec around both stars. Using the mass-temperature relations presented in Paper I, these correspond to companion masses of 0.59\Msun, 0.41\Msun, and 0.23\Msun respectively.

{\bf WASP-104.}
A companion star at $6.8\arcsec$\ was detected in both 2015 and 2016, although data were only obtained in $\rtci$ in 2015 due to a technical fault. The companion is found to be slightly too bright in our 2016 photometry, but the $\vtci$ measurement is significantly contaminated by a diffraction spike. Resolved photometry for the companion is available in 2MASS and SDSS, with the colour-spectral type relations presented by \citet{2002AJ....123.3409H} indicating a spectral type of $\sim$M6 for the companion. Using the BT-Settl model isochrones \citep{2012RSPTA.370.2765A}, we find that the SDSS $riz$ photometry is consistent with the companion being a bound companion. Further supporting this hypothesis, the URAT1 catalogue indicates common proper motion, although with large uncertainties for the motion of both stars.

{\bf WASP-108.}
Located in a crowded field, we identify two stars with colours consistent with being bound to WASP-108. Companion 10 is only $2$ mag fainter in $\rtci$, but the UCAC4, NOMAD, and PPMXL catalogues all indicate that this companion does not have common proper motion with WASP-108, and so we conclude that it is an unassociated star.

For companion 4 at $8.8\arcsec$, we measure a temperature of $2950\pm70$K in 2016, and find its flux ratios to be consistent with a bound object. However, a chance alignment is relatively likely given the crowding of the region and the faintness of this star, and further evidence is required to confirm whether these two stars are physically associated.

{\bf WASP-110.}
A bright companion to WASP-110 was noted by \citet{2014arXiv1410.3449A}, who measured a separation of $4.589\pm0.016\arcsec$\ and a magnitude difference of $2.872\pm0.012$ mag in $I_c$. We measure a similar magnitude difference in $\rtci$ of $2.284\pm0.006$ mag, and a smaller difference in $\vtci$ of $2.298\pm0.004$ mag, indicating that the companion is hotter than WASP-110 with $\Teff\simeq8400$K. Whilst not resolved in most catalogues, visual inspection of multicolour DSS and 2MASS images supports the hypothesis that the two stars have similar colours.

Various catalogues indicate that WASP-110 has a high proper motion, but with a significant scatter in the actual values, likely due to confusion between the two unresolved stars. The UCAC4 catalogue does resolve the pair, giving significantly different proper motions for the two, indicating that they are not physically associated. Our astrometry, combined with Gaia DR1 positions and the separation measured by \citet{2014arXiv1410.3449A}, indicate trends in both separation (increasing) and position angle (decreasing), in accordance with the UCAC4 proper motions. We therefore conclude that these two stars are merely a chance alignment.

{\bf WASP-111.}
We detected a bright companion $5.0\arcsec$\ from WASP-111 in both 2015 and 2016, with its colour and flux ratios being consistent with a bound companion with $\Teff=3710\pm60$K. 2MASS $JHK$ photometry and Tycho/YB6 $B$ photometry recorded in NOMAD also support this conclusion. The NOMAD catalogue indicates that the two stars have common proper motion, albeit with large uncertainties on the companion's motion. Our astrometry, combined with measurements from the 2MASS and Gaia DR1 catalogues, show no significant trends in separation or position angle since 2000. We conclude that these two stars form a bound system.

{\bf WASP-121.}
Companion 1 was detected in 2016, at a separation of $7.6\arcsec$. Its red colour is consistent with a bound companion, being $7.5$ mag fainter in $\rtci$ and $9.6$ mag fainter in $\vtci$. This star is included in the 2MASS catalogue, in which it is $\sim5$ mag fainter in $JHK$, further indicating that it is a cool object, but the 2MASS photometry is too uncertain to classify the object further. Comparing the 2MASS positions to our own astrometry, we find no large change in position angle or separation, but further astrometric observations are required to confirm whether common proper motion is present.

{\bf WASP-123.}
We detected a companion at $4.8\arcsec$\ in 2016, and derived a temperature of $3240\pm80$K, the companion being marginally consistent with a bound companion. Archival 2MASS astrometry indicates that no significant change in separation or position angle has occurred, but the 2MASS position is too uncertain to confirm common proper motion.

{\bf WASP-129.}
We find companion 2, located $5.9\arcsec$\ to the West of WASP-129, to be consistent with a bound companion, being $5.6$ mag fainter in $\rtci$ and a photometric temperature of $3360\pm90$K. 2MASS astrometry is consistent with no change in the relative position of the companion, but the proper motion of WASP-129 is too low to determine if the two stars show common proper motion.

{\bf WASP-133.}
A very close companion was detected in 2016, separated from the target star by only $2.4\arcsec$, $\sim6.7$ mag fainter. From its colour, we determine a temperature of $5000\pm400$K, indicating that it is only $\sim1000$K cooler than WASP-133. As a result, we conclude that this companion is likely to be a distant background object in a chance alignment.

\section{Orbits of binary companions} \label{sect:orbits}

For a handful of wide binaries hosting transiting hot Jupiters, historical astrometry of the stellar binary is present in the Washington Double Star Catalog (WDS). For WASP-77AB and WASP-85AB, orbital motion is clearly present, but the data cover only a small fraction of the orbit. This results in loosely constrained (and often multi-modal) orbital parameters that are difficult for Markov Chain Monte Carlo algorithms to explore. Instead, we adopt the Bayesian rejection-sampling algorithm detailed by \citet{2017AJ....153..229B}, which has been shown to converge significantly faster in such cases. For each system, the algorithm was run until 1,000,000 orbital fits had been accepted.

The WDS does not record uncertainties for the measured astrometry, and only for a few of the most recent results are uncertainties quoted in the original literature. Where we were unable to find any measurements, we assumed uncertainties of $0.1\arcsec$\ in separation and $1\degrees$ in position angle, these values being chosen by visual inspection of the scatter between measurements. 

For both systems, radial velocity measurements were obtained for both of the stellar components as part of the planet follow-up programs, in order to confirm which of the stars exhibited radial velocity variations and hence hosts a planet. Compared to the long orbital period of the binary, these measurements provide only a snapshot of the radial velocity of the system, but do allow us to break the degeneracy in the longitude of ascending node, $\Omega$, and hence constrain the full 3 dimensional motion of the system. However, without significant phase coverage it is not possible to include as a free parameter the systematic shifts in measured radial velocity caused by gravitational redshift and convective blueshift, as has been done previously in the case of $\alpha$ Cen \citep{2016A&A...586A..90P, 2017A&A...598L...7K}. These effects vary on a line-by-line basis, currently require full 3D hydrodynamical modelling to match Solar observations, and also depend on both instrumental effects and the reduction method used \citep{2013A&A...550A.103A}; given the variety of instruments and reduction pipelines used to derive the available radial velocities, we have instead opted to account for these effects with an increased uncertainty of $0.5$km/s.

We show the fitted orbits on the RA-Dec plane and show the mean, median, etc. of the resulting parameter distributions in the following two subsections. Additional tables containing the astrometric data used, figures showing the predicted separation/position angle, and figures showing the correlations between the parameters can be found in Appendix~\ref{appdx:additionalInfo}.

\subsection{WASP-77}

WASP-77AB was observed numerous times between 1903 and 1933, with the data indicating a change in position angle of $\sim2\degrees$\ and an increase in separation of $\sim0.3\arcsec$. In addition, radial velocity measurements of the individual A and B components were presented by \citet{2013PASP..125...48M}, from which we derive a radial velocity difference of $1.1\pm0.5$km/s at a mean BJD of 2455847. We constrain the total mass of the system using the stellar parameters from \citet{2013PASP..125...48M} assuming the errors on each component's mass are independent, giving $M_{\rm Tot}=1.71\pm0.07\Msun$.

\begin{figure} 
	\includegraphics[width=\columnwidth,angle=0]{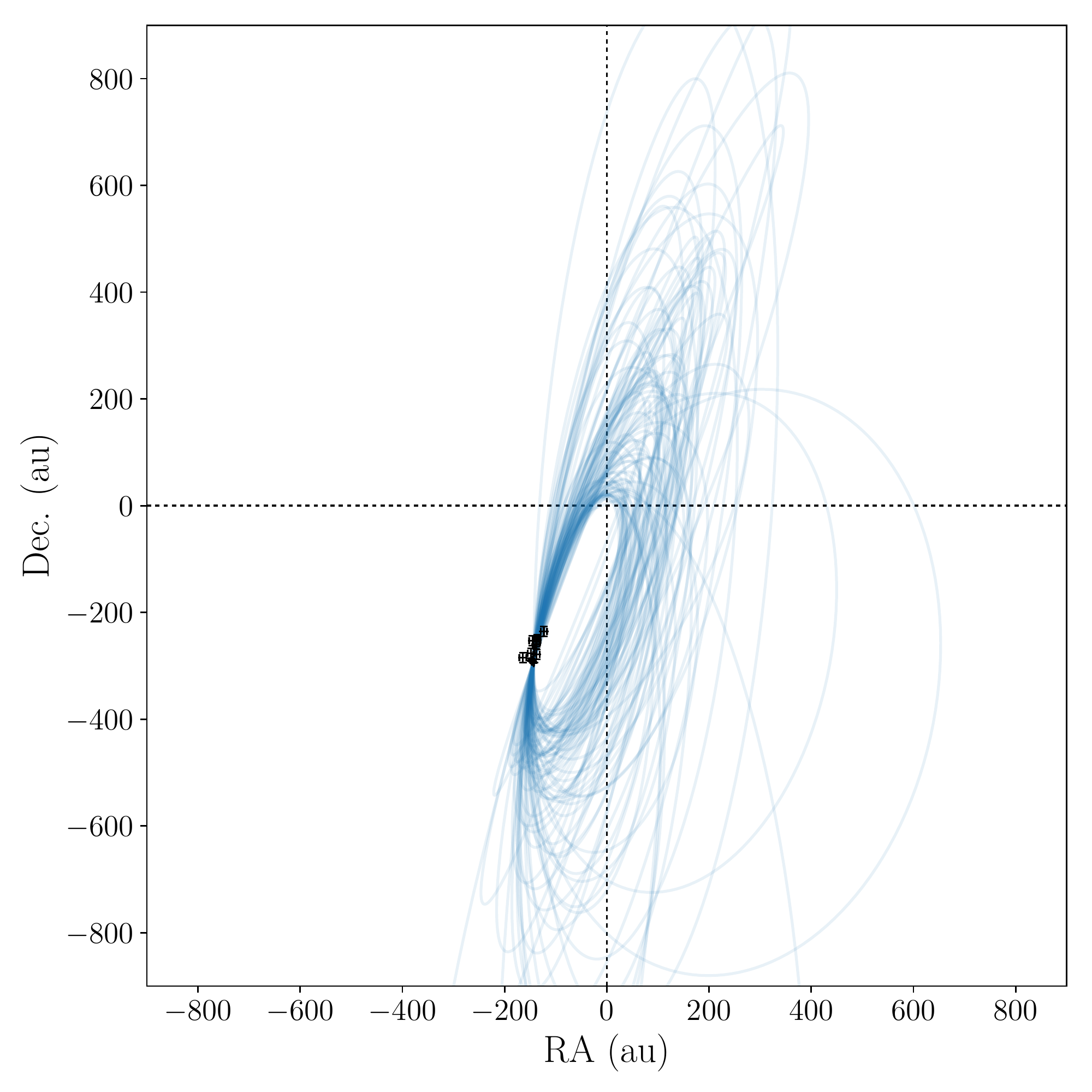}
	\caption{\label{fig:W77orbits} A selection of 100 orbits generated to fit WASP-77AB, centred on the primary star. Black data points show measured positions of the B component relative to the A component. }
\end{figure}

Our resulting parameter distributions are summarised in Table~\ref{tab:W77Orbit}. A selection of 100 orbits drawn from our accepted fits are plotted in Fig.~\ref{fig:W77orbits}. We find that the binary orbit has a semi-major axis of $461^{+200}_{-140}$au, is nearly edge-on ($i=77^{+5}_{-7}$$\degrees$), and is moderately eccentric ($e=0.51^{+0.26}_{-0.22}$). The constraint provided by the radial velocity measurements is sufficient to break the degeneracy in $\Omega$.

\begin{table*}
	\caption{\label{tab:W77Orbit} Characteristics of the probability distributions for the orbital parameters of WASP-77AB. The 16th, 50th, and 84th percentiles are denoted by $P_{16}$, $P_{50}$ and $P_{84}$. }
	\centering
	\begin{tabular}{l c c c c c c c c} \hline \hline 
		Parameter & Symbol & Median & Mean & Mode & $P_{84}$ & $P_{16}$ & $P_{84}-P_{50}$ & $P_{50}-P_{16}$ \\
		\hline
		Semi-major axis (au)   & a & 420   & 490   & 310   & 660   & 290   & 250   & 130\\
		Eccentricity$\ast$     & e & 0.60  & 0.60  & 0.95  & 0.88  & 0.34  & 0.28  & 0.26\\
		Inclination (\degrees) & i & 75    & 70    & 79    & 81    & 60    & 6     & 15 \\
		Long. of the asc. node & $\Omega$ & 339 & 313 & 340 & 346 & 299 &  7 & 40 \\
		Arg. of periapsis      & $\omega$ & 226 & 229 & 211 & 278 & 197 & 51 & 30 \\
		\hline
	\end{tabular}
	\tablefoot{$\ast$ The eccentricity distribution is bimodal, with moderate eccentricity and high eccentricity solutions ($e\simeq0.5$ and $e\simeq0.95$).}
\end{table*}

\subsection{WASP-85}

WASP-85AB has been regularly observed since 1881. Historic measurements show a large scatter in separation, but a clear trend is seen in position angle, from approximately 113\degrees\ to modern measurements near 100\degrees. \citet{2015EPSC...10..603B} obtained radial velocity measurements of the two individual stars using HARPS in good seeing, from which we derive a radial velocity difference of $-0.5\pm0.5$km/s at a mean BJD of 2456024. We constrain the total mass of the system using the stellar parameters from \citet{2015EPSC...10..603B} assuming the errors on each component's mass are independent, giving $M_{\rm Tot}=1.92\pm0.10\Msun$.

\begin{figure} 
	\includegraphics[width=\columnwidth,angle=0]{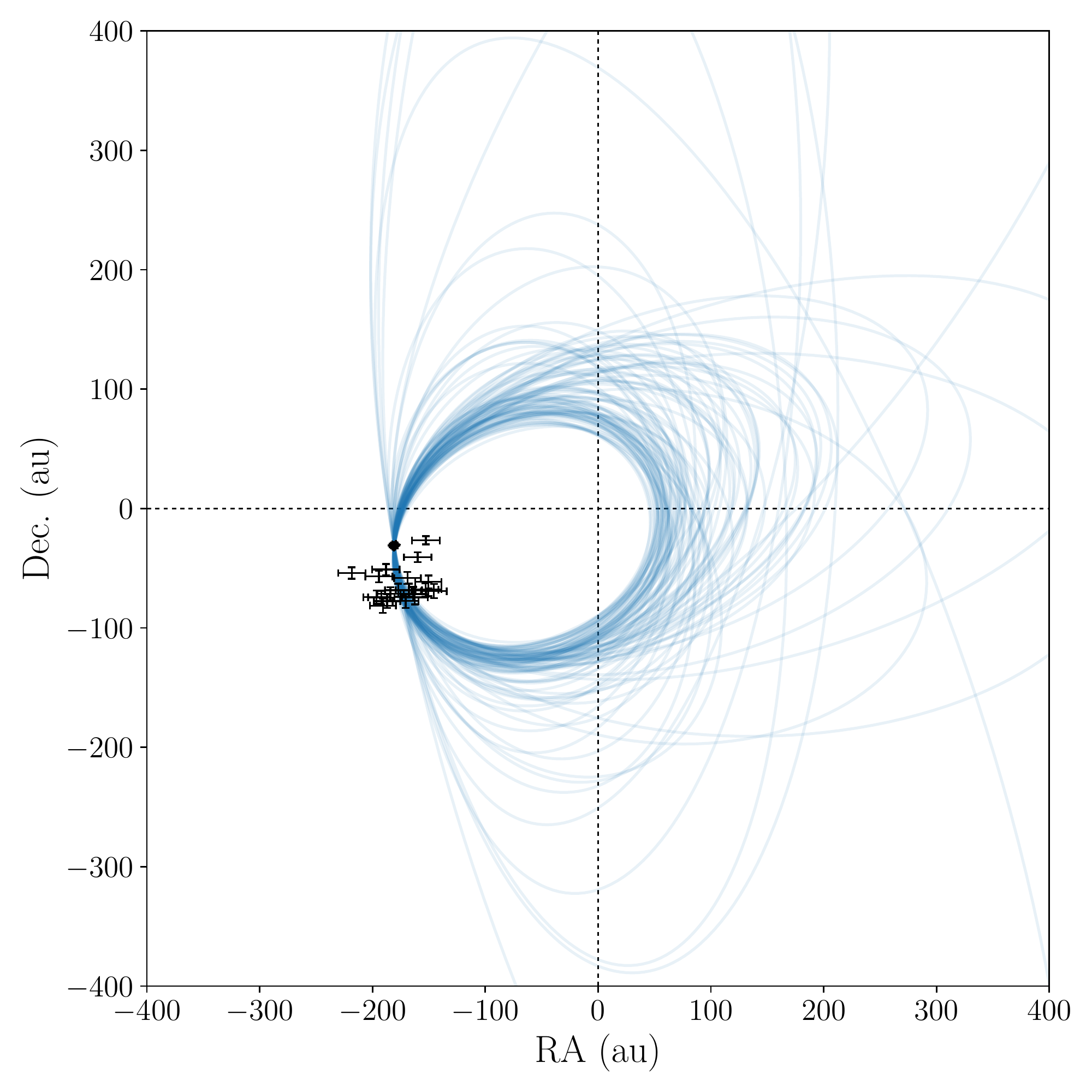}
	\caption{\label{fig:W85orbits} A selection of 100 orbits generated to fit WASP-85AB, centred on the primary star. Black data points show measured positions of the B component relative to the A component. }
\end{figure}

Our resulting parameter distributions are summarised in Table~\ref{tab:W85Orbit}. A selection of 100 orbits drawn from our accepted fits are plotted in Fig.~\ref{fig:W85orbits}. We derive $a=148^{+52}_{-23}$au, $e=0.43^{+0.13}_{-0.25}$ and $i=140^{+16}_{-12}$$\degrees$. However, we note that these parameters are strongly correlated, with smaller semi-major axes corresponding to more eccentric orbits that are closer to face on, and that the distributions are clearly non-gaussian, particularly for $a$ and $e$. The radial velocity constraint allows us to prefer one of the orbital orientations, but is not sufficient to reject the other.

When considering dynamical interactions between the planet and the outer star, one of the most important parameters is the relative inclination of the two orbits. For both the transiting planet and resolved binary, the orbital inclinations $i_{\rm pl}$ and $i_{\rm bin}$ relative to the line of sight is easily determined, and as mentioned previously, the longitude of ascending node of a visual binary orbit $\Omega_{\rm bin}$ can be determined if radial velocity measurements are available. However, $\Omega_{\rm pl}$ is entirely unconstrained, as the orientation of a transiting planet's orbit with respect to the plane of the sky is unknown unless the planet is directly imaged, which is not possible for hot Jupiters. Therefore, in order to derive the distribution of the relative inclination between the planetary and binary orbits from our data, we assume a uniform prior of $0\degrees \leq \Omega_{\rm pl} < 360\degrees$.

\begin{table*}
	\caption{\label{tab:W85Orbit} Characteristics of the probability distributions for the orbital parameters of WASP-85AB. The 16th, 50th, and 84th percentiles are denoted by $P_{16}$, $P_{50}$ and $P_{84}$. As the values of the longitude of the ascending node ($\Omega$) and argument of periapsis ($\omega$) are not fully constrained by the radial velocity measurements, we present these values both over the full range (0-360\degrees) and the restricted range usually quoted for astrometric orbits (0-180\degrees). }
	\centering
	\begin{tabular}{l c c c c c c c c} \hline \hline 
		Parameter & Symbol & Median & Mean & Mode & $P_{84}$ & $P_{16}$ & $P_{84}-P_{50}$ & $P_{50}-P_{16}$ \\
		\hline
		Semi-major axis (au)   & a & 148   & 173   & 123   & 200   & 125   & 52    & 23   \\
		Eccentricity           & e & 0.43  & 0.39  & 0.55  & 0.56  & 0.18  & 0.13  & 0.25 \\
		Inclination (\degrees) & i & 140   & 142   & 136   & 157   & 128   & 16    & 12   \\
		Long. of the asc. node & $\Omega$ & 112 & 144 & 108 & 290 & 64 & 177 & 49 \\
		Arg. of periapsis$\ast$& $\omega$ & 209 & 197 & 329 & 319 & 54 & 110 & 156 \\
		\hline
		\multicolumn{9}{c}{$\Omega$ and $\omega$ restricted to the range 0--180\degrees} \\
		Long. of the asc. node & $\Omega$ & 108 & 99 & 109 & 137 & 50 & 29 & 57 \\
		Arg. of periapsis$\ast$& $\omega$ & 81  & 86 & 36  & 154 & 24 & 73 & 57 \\
		\hline
	\end{tabular}
	\tablefoot{$\ast$ The probability distribution for $\omega$ is multimodal and is highly correlated with other parameters }
\end{table*}

\subsection{Application to planet migration theory}

The proposed eccentric Lidov-Kozai mechanism of hot Jupiter migration invokes a scenario in which an inclined wide binary companion forces an initially `cold' planet into a highly eccentric orbit, which then shrinks through tidal friction during close encounters with the planet host star (see \citealt{2016ARA&A..54..441N} for a review of the topic). One well-known prediction of this theory is that the migration will induce a spin-orbit misalignment between the planet and its host star, by changing the orientation of the planet's orbital plane. Theoretical work has also predicted the distribution of inclination angles between inner (planetary) and outer (binary) orbital planes, with numerical studies indicating that the distribution peaks for inclination angles around $40-60\degrees$ and $120-140\degrees$ \citep{2007ApJ...669.1298F, 2007ApJ...670..820W, 2012ApJ...754L..36N}.

For WASP-77, the projected inclination between the two orbital planes is $22\pm16\degrees$, and for the WASP-85 system it is $60\pm17\degrees$, adopting the planetary orbital inclinations from \citet{2013PASP..125...48M} and \citet{2016AJ....151..150M} respectively. Determining the true inclination between the orbits is complicated by our lack of knowledge regarding the true orientation of the planetary orbit, as it is impossible to determine $\Omega_p$ without astrometric measurements of the orbit or direct imaging of the planet. We estimate the true inclination angle using a Monte Carlo simulation, choosing $10^6$ values for $\Omega_p$ from a uniform distribution ($0\degrees \leq \Omega_p < 360\degrees$), with our resulting 1-$\sigma$ confidence limits being $36\degrees<i<144\degrees$ for WASP-77 and $59\degrees<i<121\degrees$ for WASP-85.

We briefly note that many analyses of the Lidov-Kozai effect assume a circular orbit for the outer (stellar) binary. However, it has been shown that significantly different behaviour can result from eccentric outer orbits \citep{2011Natur.473..187N, 2013MNRAS.431.2155N} -- in particular, the Lidov-Kozai effect is able to occur in many more orbital configurations for eccentric systems than for circular. Our fitted binary orbits do indeed have non-negligible eccentricity, which is in line with results from studies of field stars, which show that many wide stellar binaries are eccentric \citep{2010ApJS..190....1R}

\section{Discussion}

We find good agreement with other high-resolution imaging studies that also observed our targets, with Paper I discussing a small number of faint or close-in companions that were discovered by others, but fell below our detection limits. Since Paper I, we also reported on the discovery of a close-in companion to WASP-20 at a separation of $0.26\arcsec$, which fell below the sensitivity limit of our lucky imaging observations \citep{2016ApJ...833L..19E}. For the targets newly added to our survey since Paper I, no directly imaged companions went undetected in our data. However, this is at least partly due to the fact that no other survey has yet covered many of these these recently announced TEP systems, with the recent work of \citet{2016ApJ...827....8N} mainly covering systems announced before 2013.

A small number of new targets in our survey have direct imaging observations reported in an individual basis, with some AO or lucky imaging observations being performed during the follow-up of KELT and HAT-South targets, whilst K2 targets are regularly inspected with adaptive optics imaging. Our observations have detected a handful of very faint companions to K2 targets that were at or below the quoted sensitivity limits of the AO observations, with the results otherwise being comparable; comments on individual systems are given in the previous section.

\subsection{The lack of bright companions to hot Jupiters}

In the reduction of our data, it was noted that among the companions that were likely to be bound, there is a deficit of bright stars compared to faint ones: of the 45 such companions listed in Table~\ref{tab:comps2sigma}, none are brighter than the planet host star, only two are less than one mag fainter ($4\pm3\%$), and only five are less than two mag fainter ($11\pm5\%$). Across our range of target stars, a binary companion 2 mag fainter in \rtci approximately corresponds to a binary mass ratio $q$ of 0.75, and so $11\pm5\%$ of hot Jupiter binaries detected have $q>0.75$. This contrasts with the results of recent surveys of field stars, where the distribution of mass ratios (and hence brightness ratios) shows no preference towards low values, with the results of \citet{2010ApJS..190....1R} finding that $31\pm5\%$ of field binaries have $q>0.75$. If this deficit of bright, high-mass companions does indeed exist in the population of hot Jupiter hosting systems, it would indicate that the formation of these close-in planets is inhibited in such binaries systems. This may be due to the suppression of planet formation in protoplanetary discs (e.g. \citealt{2015ApJ...799..147J}), or instead due to their influence on planetary migration.

This deficit was also seen in the Friends of Hot Jupiters survey (FoHJ, \citealt{2015ApJ...800..138N, 2016ApJ...827....8N}), with all but 2 of their 35 candidate binary companions being at least 2.5 mag fainter in the K band, but the opposite result was found by \citet{2015ApJ...806..248W} in a sample of hot Jupiters from the Kepler mission, with six out of their seven detected companion stars being within 2.5 mag of the planet host star. \citet{2015ApJ...806..248W} hypothesised that their high fraction of bright companions could be a result of the migration mechanism of hot Jupiters, with higher mass stellar companions being more easily able to migrate planets inwards via the Kozai-Lidov mechanism. The lack of bright companions to hot Jupiters discovered by ground-based planet surveys was attributed to selection biases, with the detection of such planets being inhibited by photometric dilution, combined with cautious follow-up strategies that avoid systems showing any potential evidence of binary nature, such as multiple sets of spectral lines \citep{2017arXiv170707521T}. However, due to the qualitative nature of target selection, there are no detailed assessments of the completeness of ground-based planet surveys, and hence we cannot assess the likely number of hot Jupiters in wide stellar binaries that have been missed by such surveys.

On the other hand, even if the majority of hot Jupiters do live in high mass binaries as suggested by \citet{2015ApJ...806..248W}, there is still a notable population of hot Jupiters for which no stellar companion has been identified. The FoHJ survey found that $49\%$\footnote{Combining their multiplicity rates of $3.9^{+4.5}_{-2.0}\%$ and $47\pm7$\% within and beyond 50 au, respectively.} of systems in their sample have no binary stellar companion \citep{2016ApJ...827....8N}. This suggests that some hot Jupiters formed in stellar binaries that have since been disrupted, leaving only a single star, or alternatively that multiple hot Jupiter formation mechanisms operate, as has been proposed by \citet{2017arXiv170309711N}.

\subsection{Companion detection using Gaia}

Future detection and confirmation of well-separated binary companions will be made significantly easier with Gaia data. Companions of moderate brightness ($G\lessapprox18$) at typical distances of a few hundred parsecs are expected to have parallaxes measured to better than 10\% accuracy \citep{2014EAS....67...23D}, sufficient to rule out the vast majority of background stars present in our images. Proper motions for even the faintest companions are expected to have errors less than $1$ mas/yr, based on the expected parallax-proper motion error relations \citep{2016A&A...595A...1G}, sufficient to show common proper motion for all but the slowest moving planet host stars.

Gaia's ability to directly resolve close companions falls below separations of a couple of arcseconds, especially for faint companions; however, the expected limits are similar to the sensitivity achieved by our ground-based lucky imaging survey. Gaia will reliably resolve equal-brightness components at separations greater than $0.6\arcsec$ ($95\%$ completeness), but successful detections of companions 6-7 magnitudes fainter will only occur in approximately $5\%$ of cases at the same separation \citep{2015A&A...576A..74D}.

Astrometric detections of orbital motion by Gaia will allow some unresolved stellar companions to be detected, with the limits from exoplanet detection studies suggesting that pairs of solar-mass stars could be detected out to orbits of a few tens of au \citep{2010EAS....42...55S} for systems within $200$pc. Combined with resolved detections of companions, Gaia should provide good coverage for bright companions to the closest transiting exoplanet systems, where the detection limits of the two methods overlap. However, the detection of faint, close-in binary companions will remain possible only with techniques such as adaptive optics and interferometry -- for example, none of the companions detected by \citet{2016AJ....152....8K} using the non-redundant aperture masking technique are expected to be directly resolved by Gaia, and the majority are thought to be in long period orbits that will not produce a detectable astrometric signal.

\section{Conclusions}

We present lucky imaging observations of 97 transiting planet systems. We analyse these data to create a catalogue of stars within the vicinity of our targets, providing relative astrometry and photometry for each companion. Where two colour photometry was possible, we estimate the stellar temperature of each detected star, and hence determine whether it is consistent with being a bound main sequence companion to the planetary system. For viable candidates, we review evidence available in the literature and public catalogues to either support or refute the `bound companion' hypothesis. Combining our data with historical astrometry, we fit the orbits of the wide stellar binaries WASP-77AB and WASP-85AB, and find both binaries to be moderately eccentric with orbital planes that are at least slightly inclined relative to the planetary orbital plane.

We consider the mass ratio distribution of our plausible binary star systems, in addition to the sample of confirmed binaries from \citet{2015ApJ...800..138N, 2016ApJ...827....8N}, and find that our targets very rarely have high mass stellar companions ($q>0.75$). Due to the complex selection biases present in ground-based planet surveys, it is difficult to determine if our target sample is a true representation of all hot Jupiters. However, the significant proportion of hot Jupiters with no detected stellar companion suggests that either formation via the binary Lidov-Kozai mechanism does not occur in all hot Jupiter systems, or instead that many stellar binaries were catastrophically disrupted at an earlier stage, leaving the hot Jupiter in a non-multiple stellar system.

\begin{longtab}
\begin{landscape}
\begin{longtable}{l l r@{\,$\pm$\,}l r@{\,$\pm$\,}l r@{\,$\pm$\,}l r@{\,$\pm$\,}l r@{\,$\pm$\,}l r@{\,$\pm$\,}l r@{\,$\pm$\,}l l l}

		\caption{\label{tab:data15} Astrometry, photometry, and derived values for our 2015 observations. The column headed $\sigma_{\rm dist}$ gives the probability that the companion is at the same distance as the associated exoplanet host star, which is calculated by comparing the observed flux ratio to the flux ratio expected if the two stars were bound, based on their temperatures. Our flux ratio prediction is based on the assumption that both stars are on the main sequence.  The full version of table is available in electronic form at the CDS. } \\
		\hline \hline 
		Name & ID &  \multicolumn{2}{c}{Separation} & \multicolumn{2}{c}{Position Angle} & \multicolumn{2}{c}{Proj. Sep} & \multicolumn{2}{c}{\dvtci} & \multicolumn{2}{c}{\drtci} & \multicolumn{2}{c}{$\vtci-\rtci$} & \multicolumn{2}{c}{\Teff} & $\sigma_{\rm dist}$ & \\
		\hline
		\endfirsthead
		\caption{continued.}\\
		Name & ID &  \multicolumn{2}{c}{Separation} & \multicolumn{2}{c}{Position Angle} & \multicolumn{2}{c}{Proj. Sep} & \multicolumn{2}{c}{\dvtci} & \multicolumn{2}{c}{\drtci} & \multicolumn{2}{c}{$\vtci-\rtci$} & \multicolumn{2}{c}{\Teff} & $\sigma_{\rm dist}$ & \\
		\hline\hline
		\hline
		\endhead
		\hline
		\endfoot
		CoRoT-22   &  1 &  3.1307 & 0.0231 &   2.861 & 0.744 &  2019 &   205 &  +5.82 & 0.09 &  +5.52 & 0.06 & +0.51 & 0.12 &  4545 &  246 &  -6.4 \\
		CoRoT-22   &  2 &  3.5276 & 0.0005 & 198.959 & 0.061 &  2275 &   230 &  +9.30 & 0.41 &  +8.26 & 0.16 & +1.25 & 0.44 &  3590 &  403 &  -2.0 \\
		CoRoT-22   &  3 &  5.1940 & 0.0082 & 210.564 & 0.107 &  3350 &   339 &  +4.84 & 0.07 &  +4.03 & 0.05 & +1.02 & 0.11 &  3805 &  119 &  -1.9 \\
		CoRoT-22   &  4 &  5.2043 & 0.0001 & 210.608 & 0.060 &  3357 &   340 &      - &    - &  +4.20 & 0.08 &     - &    - &     - &    - &     - \\
		\multicolumn{17}{c}{$\cdots$} \\
\end{longtable}
\end{landscape}
\end{longtab}

\begin{longtab}
\begin{landscape}
\begin{longtable}{l l r@{\,$\pm$\,}l r@{\,$\pm$\,}l r@{\,$\pm$\,}l r@{\,$\pm$\,}l r@{\,$\pm$\,}l r@{\,$\pm$\,}l r@{\,$\pm$\,}l l l}
			
		\caption{\label{tab:data16} Astrometry, photometry, and derived values for our 2016 observations. The column headed $\sigma_{\rm dist}$ gives the probability that the companion is at the same distance as the associated exoplanet host star, which is calculated by comparing the observed flux ratio to the flux ratio expected if the two stars were bound, based on their temperatures. Our flux ratio prediction is based on the assumption that both stars are on the main sequence.  The full version of table is available in electronic form at the CDS. } \\
		\hline \hline 
		Name & ID &  \multicolumn{2}{c}{Separation} & \multicolumn{2}{c}{Position Angle} & \multicolumn{2}{c}{Proj. Sep} & \multicolumn{2}{c}{\dvtci} & \multicolumn{2}{c}{\drtci} & \multicolumn{2}{c}{$\vtci-\rtci$} & \multicolumn{2}{c}{\Teff} & $\sigma_{\rm dist}$ & \\
		\hline
		\endfirsthead
		\caption{continued.}\\
		Name & ID &  \multicolumn{2}{c}{Separation} & \multicolumn{2}{c}{Position Angle} & \multicolumn{2}{c}{Proj. Sep} & \multicolumn{2}{c}{\dvtci} & \multicolumn{2}{c}{\drtci} & \multicolumn{2}{c}{$\vtci-\rtci$} & \multicolumn{2}{c}{\Teff} & $\sigma_{\rm dist}$ & \\
		\hline\hline
		\hline
		\endhead
		\hline
		\endfoot
		CoRoT-02   &  1 &  4.0691 & 0.0012 & 208.614 & 0.065 &   921 &    93 &  +3.73 & 0.06 &  +2.85 & 0.04 & +1.17 & 0.10 &  3657 &   91 &  +0.2 \\
		CoRoT-02   &  2 &  6.3572 & 0.0093 &  58.607 & 0.385 &  1439 &   146 & +10.00 & 0.30 &  +9.55 & 0.17 & +0.74 & 0.35 &  4161 &  531 &  -4.8 \\
		CoRoT-02   &  3 &  7.6518 & 0.0127 & 309.856 & 0.114 &  1732 &   176 &  +7.85 & 0.10 &  +7.52 & 0.06 & +0.62 & 0.13 &  4335 &  210 &  -7.8 \\
		CoRoT-02   &  4 &  9.8063 & 0.0471 &  27.601 & 0.152 &  2219 &   225 &  +9.53 & 0.25 &  +9.48 & 0.16 & +0.34 & 0.30 &  4929 &  882 &  -5.2 \\
		\multicolumn{17}{c}{$\cdots$} \\
\end{longtable}
\end{landscape}
\end{longtab}


\begin{acknowledgements}
	D.F.E. is funded by the UK’s Science and Technology Facilities Council. J. Southworth acknowledges support from the Leverhulme Trust in the form of a Philip Leverhulme prize. CvE acknowledges funding for the Stellar Astrophysics Centre, provided by The Danish National Research Foundation (Grant DNRF106). We acknowledge the use of the NASA Astrophysics Data System; the SIMBAD database and the VizieR catalogue access tool operated at CDS, Strasbourg, France; and the arXiv scientific paper preprint service operated by Cornell University. This publication makes use of data products from the Two Micron All Sky Survey, which is a joint project of the University of Massachusetts and the Infrared Processing and Analysis Center/California Institute of Technology, funded by the National Aeronautics and Space Administration and the National Science Foundation. This work has made use of data from the European Space Agency (ESA) mission {\bf } (\url{http://www.cosmos.esa.int/gaia}), processed by the {\bf Gaia} Data Processing and Analysis Consortium (DPAC, \url{http://www.cosmos.esa.int/web/gaia/dpac/consortium}). Funding for the DPAC has been provided by national institutions, in particular the institutions participating in the {\bf Gaia} Multilateral Agreement. This research has made use of the Washington Double Star Catalog maintained at the U.S. Naval Observatory. We thank Sarah Blunt, Eric Nielsen, and Robert De Rosa for helpful comments and advice on the use of their OFTI algorithm.
\end{acknowledgements}


\bibliographystyle{aa}
\bibliography{dfe}


\begin{appendix} 

	\section{Additional tables and figures relating to the orbits of WASP-77AB and WASP-85AB} \label{appdx:additionalInfo}
	
\begin{figure*}
			\includegraphics[width=\textwidth,angle=0]{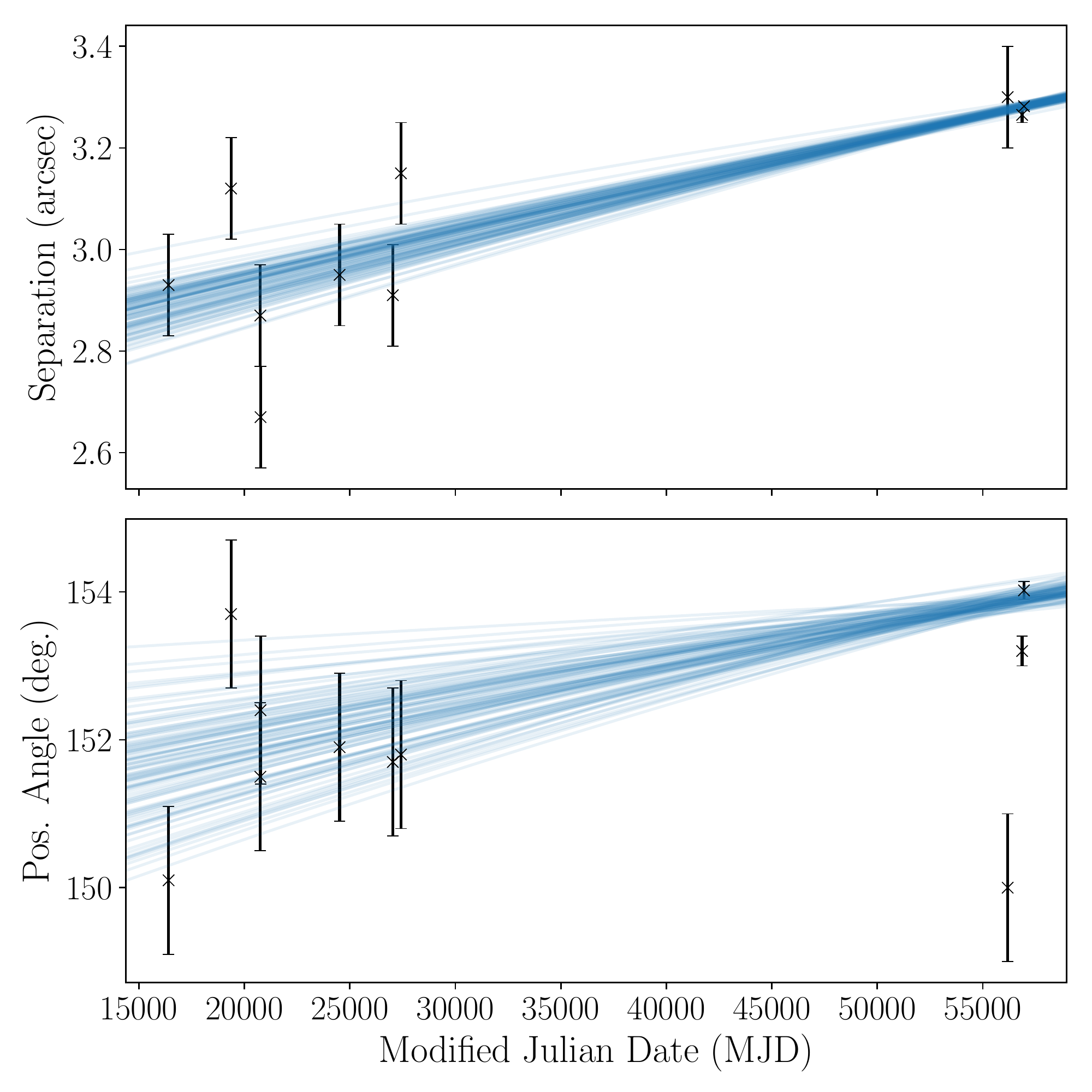}
			\caption{\label{fig:W77FitvsObs} A comparison between the measured separations and position angles of WASP-77AB, and those predicted by the 100 randomly chosen models presented in Fig.~\ref{fig:W77orbits}. }
		\end{figure*}

\begin{figure*} 
			\includegraphics[width=\textwidth,angle=0]{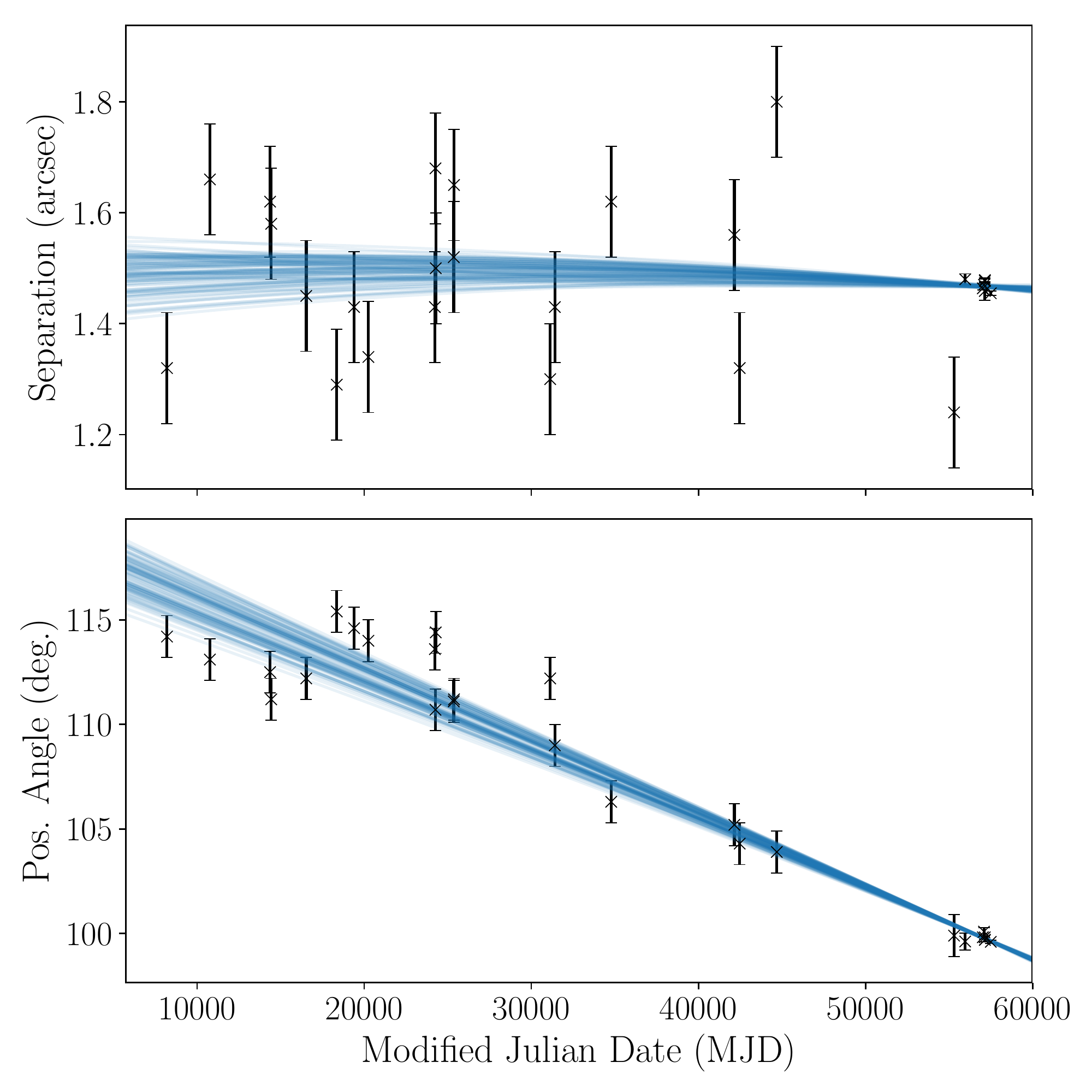}
			\caption{\label{fig:W85FitvsObs} A comparison between the measured separations and position angles of WASP-85AB, and those predicted by the 100 randomly chosen models presented in Fig.~\ref{fig:W85orbits}. }
		\end{figure*}

\begin{figure*}
		\includegraphics[width=\textwidth,angle=0]{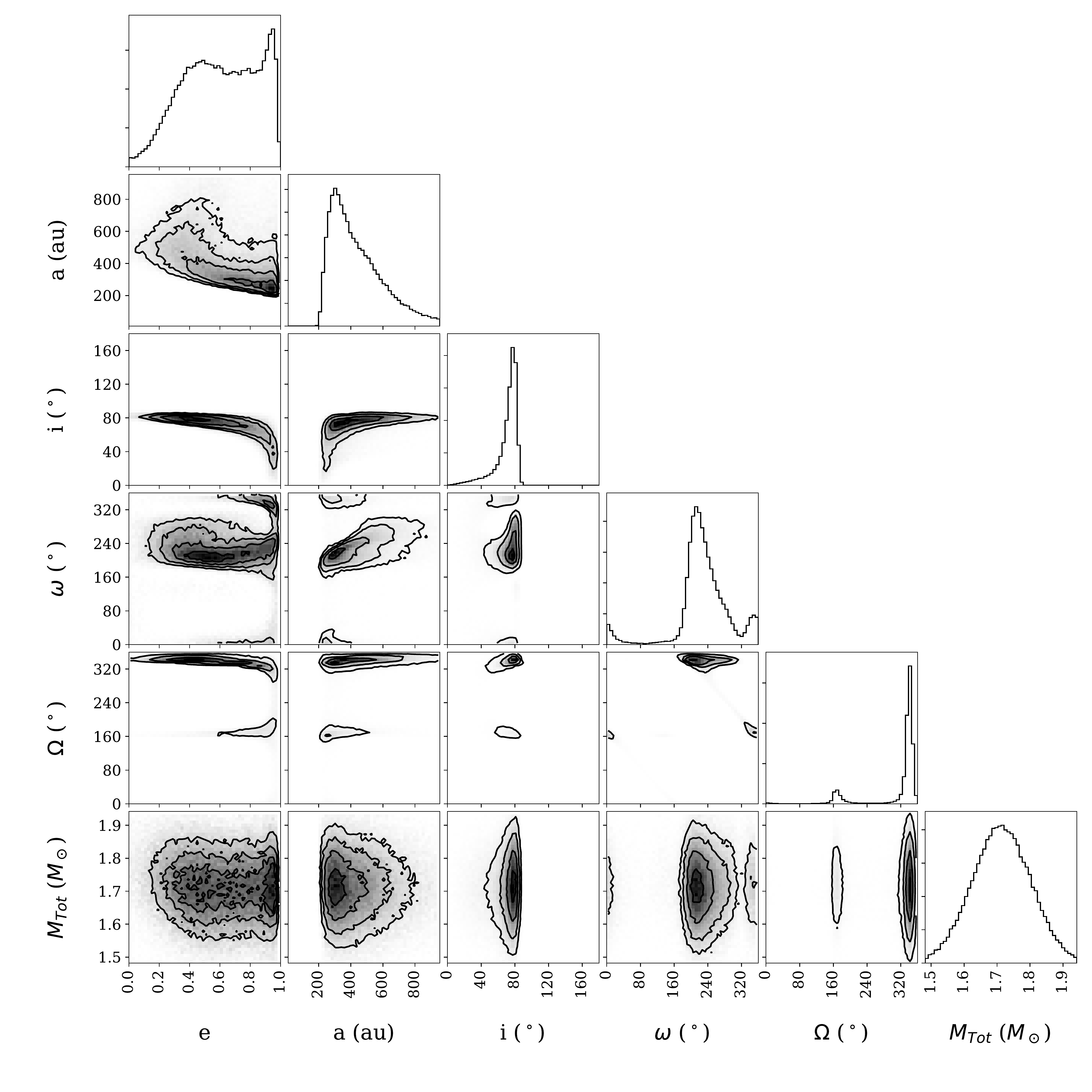}
		\caption{\label{fig:W77corner} Density distributions of the parameters of the 1,000,000 accepted fits to the WASP-77AB binary orbit, based on the model described in Sect.~\ref{sect:orbits}. The consecutive contours encompass 68.3\%, 95.5\%, 99.7\% and 99.9\% of the samples. The symbols used are defined as: $a$ -- semi-major axis; $e$ -- eccentricity; $i$ -- inclination; $\omega$ -- longitude of the ascending node; $\Omega$ -- argument of periapsis; $M_{\rm Tot}$ -- total mass of the system.}
	\end{figure*}

\begin{figure*} 
		\includegraphics[width=\textwidth,angle=0]{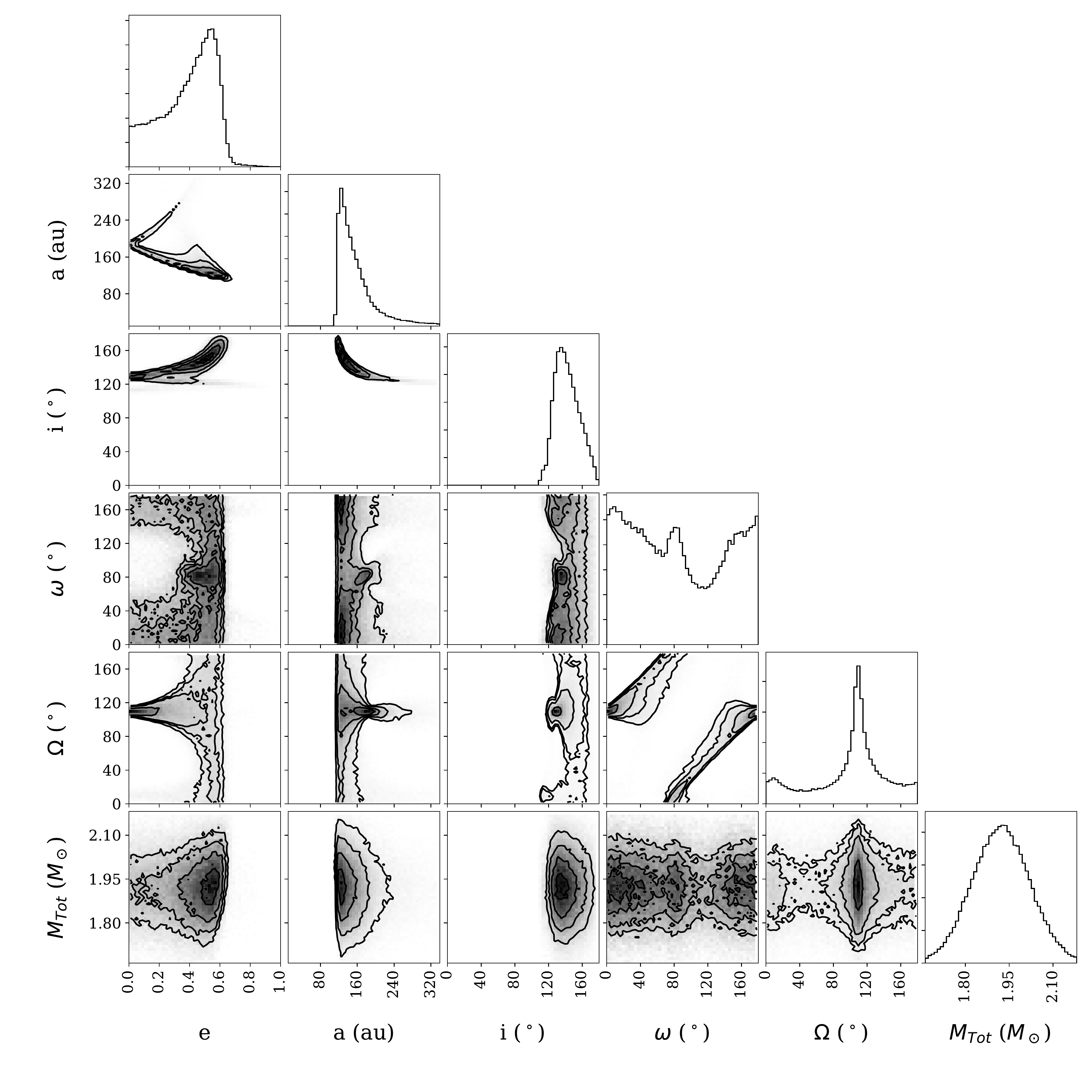}
		\caption{\label{fig:W85corner} Density distributions of the parameters of the 1,000,000 accepted fits to the WASP-85AB binary orbit, based on the model described in Sect.~\ref{sect:orbits}. The consecutive contours encompass 68.3\%, 95.5\%, 99.7\% and 99.9\% of the samples. The symbols used are defined as: $a$ -- semi-major axis; $e$ -- eccentricity; $i$ -- inclination; $\omega$ -- longitude of the ascending node; $\Omega$ -- argument of periapsis; $M_{\rm Tot}$ -- total mass of the system. Due to a lack of a strong radial velocity constraint on $\omega$ and $\Omega$, we have limited these two values to the range $0 \leq x < 180$. }
	\end{figure*}

	This appendix contains supporting information on the orbital fitting process and its results, described in Section~\ref{sect:orbits}. Tables~\ref{tab:W77astrometry}~and~\ref{tab:W85astrometry} list the astrometric data used to fit the orbits, mainly taken from the Washington Double Star Catalogue (WDS). As discussed in Section~\ref{sect:orbits}, uncertainties of $0.1\arcsec$\ in separation and $1\degrees$ in position angle were assumed where no values were quoted in the literature; these points contain `WDS' in the reference column. Figures~\ref{fig:W77FitvsObs}~and~\ref{fig:W85FitvsObs} compare the measured separations and position angles to those predicted by 100 randomly chosen models (these are same 100 models plotted in Figures~\ref{fig:W77orbits}~and~\ref{fig:W85orbits}). Finally, figures~\ref{fig:W77corner}~and~\ref{fig:W85corner} show the distribution of orbital parameters given by the fitting process described in Section~\ref{sect:orbits}.
	
	\begin{table*}
		\caption{\label{tab:W77astrometry} Available astrometric data for the WASP-77AB system. Data with `WDS' in the reference column are taken from the Washington Double Star catalog, which contains further information on the provenance of these data.}
		\centering
		\begin{tabular}{l r@{\,$\pm$\,}l r@{\,$\pm$\,}l l} \hline \hline 
			Date (MJD) & \multicolumn{2}{c}{Separation ($\arcsec$)} & \multicolumn{2}{c}{Position Angle ($\degrees$)} & Reference \\
			\hline
			16411 & 2.93 & 0.10 & 150.1 & 1.0 & WDS \\
			19370 & 3.12 & 0.10 & 153.7 & 1.0 & WDS \\
			20761 & 2.87 & 0.10 & 151.5 & 1.0 & WDS \\
			20769 & 2.67 & 0.10 & 152.4 & 1.0 & WDS \\
			24516 & 2.95 & 0.10 & 151.9 & 1.0 & WDS \\
			27040 & 2.91 & 0.10 & 151.8 & 1.0 & WDS \\
			27423 & 3.15 & 0.10 & 151.8 & 1.0 & WDS \\
			56180 & 3.3 & 0.10 & 150.0 & 1.0 & 1 \\
			56863 & 3.265 & 0.015 & 153.2 & 0.2 & 2 \\
			56952 & 3.282 & 0.007 & 154.02 & 0.12 & 3 \\
			\hline	
		\end{tabular}
		\tablebib{(1) \citet{2013PASP..125...48M}; (2) Paper I \citep{HITEP1}; (3) \citet{2015A&A...575A..23W}.}
	\end{table*}
	
	\begin{table*}
		\caption{\label{tab:W85astrometry} Available astrometric data for the WASP-85AB system. Data with `WDS' in the reference column are taken from the Washington Double Star catalog, which contains further information on the provenance of these data.}
		\centering
		\begin{tabular}{l r@{\,$\pm$\,}l r@{\,$\pm$\,}l l} \hline \hline 
			Date (MJD) & \multicolumn{2}{c}{Separation ($\arcsec$)} & \multicolumn{2}{c}{Position Angle ($\degrees$)} & Reference \\
				\hline
				\hphantom{5}8197 & 1.32 & 0.10 & 114.2 & 1.0 & WDS \\
				10765 & 1.66 & 0.10 & 113.1 & 1.0 & WDS \\
				14368 & 1.62 & 0.10 & 112.5 & 1.0 & WDS \\
				14435 & 1.58 & 0.10 & 111.2 & 1.0 & WDS \\
				16531 & 1.45 & 0.10 & 112.2 & 1.0 & WDS \\
				18358 & 1.29 & 0.10 & 115.4 & 1.0 & WDS \\
				19395 & 1.43 & 0.10 & 114.6 & 1.0 & WDS \\
				20250 & 1.34 & 0.10 & 114.0 & 1.0 & WDS \\
				24231 & 1.43 & 0.10 & 113.6 & 1.0 & WDS \\
				24264 & 1.68 & 0.10 & 110.7 & 1.0 & WDS \\
				24282 & 1.50 & 0.10 & 114.4 & 1.0 & WDS \\
				25360 & 1.52 & 0.10 & 111.1 & 1.0 & WDS \\
				25374 & 1.65 & 0.10 & 111.2 & 1.0 & WDS \\
				31131 & 1.30 & 0.10 & 112.2 & 1.0 & WDS \\
				31408 & 1.43 & 0.10 & 109.0 & 1.0 & WDS \\
				34783 & 1.62 & 0.10 & 106.3 & 1.0 & WDS \\
				42152 & 1.56 & 0.10 & 105.2 & 1.0 & WDS \\
				42472 & 1.32 & 0.10 & 104.3 & 1.0 & WDS \\
				44692 & 1.80 & 0.10 & 103.9 & 1.0 & WDS \\
				55302 & 1.24 & 0.10 & 99.9 & 1.0 & WDS \\
				55967 & 1.48 & 0.01 & 99.62 & 0.41 & 1 \\
				57023 & 1.4637 & 0.003 & 99.824 & 0.016 & 2 \\
				57091 & 1.470 & 0.003 & 100.09 & 0.19 & 3 \\
				57141 & 1.4590 & 0.017 & 99.819 & 0.084 & This work \\
				57146$^\ast$ & 1.4769 & 0.004 & 99.7 & 0.1 & 4 \\
				57146$^\ast$ & 1.4786 & 0.004 & 99.7 & 0.1 & 5 \\
				57499 & 1.4548 & 0.0038 & 99.606 & 0.083 & This work \\
				\hline	
			\end{tabular}
			\tablefoot{$^\ast$It appears that the data presented in references 4 and 5 are separate reductions of the same observations. We treat them as independent measurements.}
			\tablebib{(1) \citet{2015EPSC...10..603B}; (2) Gaia DR1 \citep{2016A&A...595A...1G}; (3) \citet{2015A&A...579A.129W} (4) \citet{2016AJ....151..153T}; (5) \citet{2016AJ....151..159S};}
		\end{table*}
		
\end{appendix}

\end{document}